\documentclass[twocolumn,preprintnumbers,amsmath,amssymb,pra,floatfix,superscriptaddress]{revtex4-2}
\usepackage{graphicx}
\usepackage{dcolumn}
\usepackage{bm}
\usepackage{color}
\usepackage{braket}
\usepackage{float}
\usepackage[caption=false]{subfig}
\usepackage{nicefrac}
\usepackage{dcolumn}

\begin{document}

\title{Two-photon photoassociation spectroscopy of the $^{2}\Sigma^+$ YbLi molecular ground state}

\author{Alaina Green}
\email{agreen13@uw.edu}
\affiliation{Department of Physics, University of Washington, Seattle, Washington 98195, USA}
\author{Jun Hui See Toh}
\affiliation{Department of Physics, University of Washington, Seattle, Washington 98195, USA}
\author{ Richard Roy}
\affiliation{Department of Physics, University of Washington, Seattle, Washington 98195, USA}
\author{ Ming Li}
\affiliation{Department of Physics, Temple University, Philadelphia, Pennsylvania 19122, USA}
\author{Svetlana Kotochigova}
\affiliation{Department of Physics, Temple University, Philadelphia, Pennsylvania 19122, USA}
\author{ Subhadeep Gupta}
\affiliation{Department of Physics, University of Washington, Seattle, Washington 98195, USA}

\date{\today}
\begin{abstract}

We report on measurements of the binding energies of several weakly bound vibrational states of the paramagnetic $^{174}$Yb$^{6}$Li molecule in the electronic ground state using two-photon spectroscopy in an ultracold atomic mixture confined in an optical dipole trap. We theoretically analyze the experimental spectrum to obtain an accurate description of the long-range potential of the ground state molecule. Based on the measured binding energies, we arrive at an improved value of the interspecies $s$-wave scattering length $a_{s0}=30$ $a_0$. Employing coherent two-photon spectroscopy we also observe the creation of ``dark'' atom-molecule superposition states in the heteronuclear Yb-Li system. This work is an important step towards the efficient production of ultracold YbLi molecules via association from an ultracold atomic mixture.

\end{abstract}
\maketitle

\section{Introduction}

Ultracold molecules expand the scientific reach of ultracold atoms through their richer internal structure. Samples of ultracold polar molecules are sought for their tunable long-range interactions which form the basis for studies of strongly correlated many-body systems as well as serving as building blocks for quantum information processing~\cite{carr09,gadw16}. Precision spectroscopies on ultracold molecules can be used to search for time variations of fundamental constants~\cite{zele08,huds08,kaji08} and test fundamental symmetries~\cite{flaum14,huds11,ande18}. To reach the high phase space densities (PSD) required for several of these proposed applications of ultracold molecules, an established strategy is to coherently transform pairs of atoms from within high PSD atomic mixtures into a sample of ultracold ground state molecules through a sequence of magneto-association and Stimulated Raman Adiabatic Passage (StiRAP) processes~\cite{ni08,take14,molo14,park15,Guo16,rvac17}. 

While this strategy has been successful so far only in bi-alkali atom pairs, there is general interest in extending it to other diatomic systems such as dimers containing one alkaline-earth-like and one alkali atom (e.g. YbLi), which are attractive due to the new degree of freedom from the unpaired electron. While this difference in electronic structure make magneto-association more challenging in these systems ~\cite{brue12, dowd15, barb18}, several groups are actively pursuing this experimental front \cite{munc11,roy16PA,ciam18,gutt18}. The resulting combination of electric and magnetic dipole moments in the molecular ground state are appealing properties for their proposed use towards quantum simulation of spin lattice models and studies of symmetry-protected topological phases, spin liquids, and quantum magnetism ~\cite{mich06,gadw16}. The $^{2}\Sigma^+$ ground state of such molecules also make them promising candidates for fundamental symmetry tests as well as ultracold chemistry with a spin degree of freedom~\cite{carr09,gadw16}. 

In this paper, we report on first measurements probing bound states of the $^{174}$Yb$^{6}$Li $^{2}\Sigma^+$ ground state molecule by two-photon photoassociation (PA) spectroscopy. We observe several weakly bound states constituting a single vibrational spectrum, which we theoretically analyze to accurately determine the long-range dispersion coefficients for this potential. Our analysis also yields an improved value of the YbLi ground state $s$-wave scattering length. Through coherent two-photon spectroscopy, we also observe ``dark'' atom-molecule superposition states and perform a precise measurement of the least-bound vibrational state energy. Our results are an important step towards efficient coherent production of YbLi molecules via association from ultracold atoms. 

The rest of this paper is organized in the following way. In Section~\ref{sec:twophspec}, we describe our experimental method and results for the determination of vibrational states in YbLi using PA spectroscopy. In Section~\ref{sec:theory}, we describe our theoretical analysis of the observed spectrum and determination of the long-range molecular potential, and in Section~\ref{sec:scatlength} we discuss the resulting value for the ground state $s$-wave scattering length. We present our observation of atom-molecule superposition states in Section~\ref{sec:darkstatespec} and provide a summary and outlook in Section~\ref{sec:summary}.

\section{\label{sec:twophspec} Two-photon Spectroscopy} 

\subsection{Experimental strategy}

\begin{figure}
\begin{centering}
\includegraphics[width=1\columnwidth]{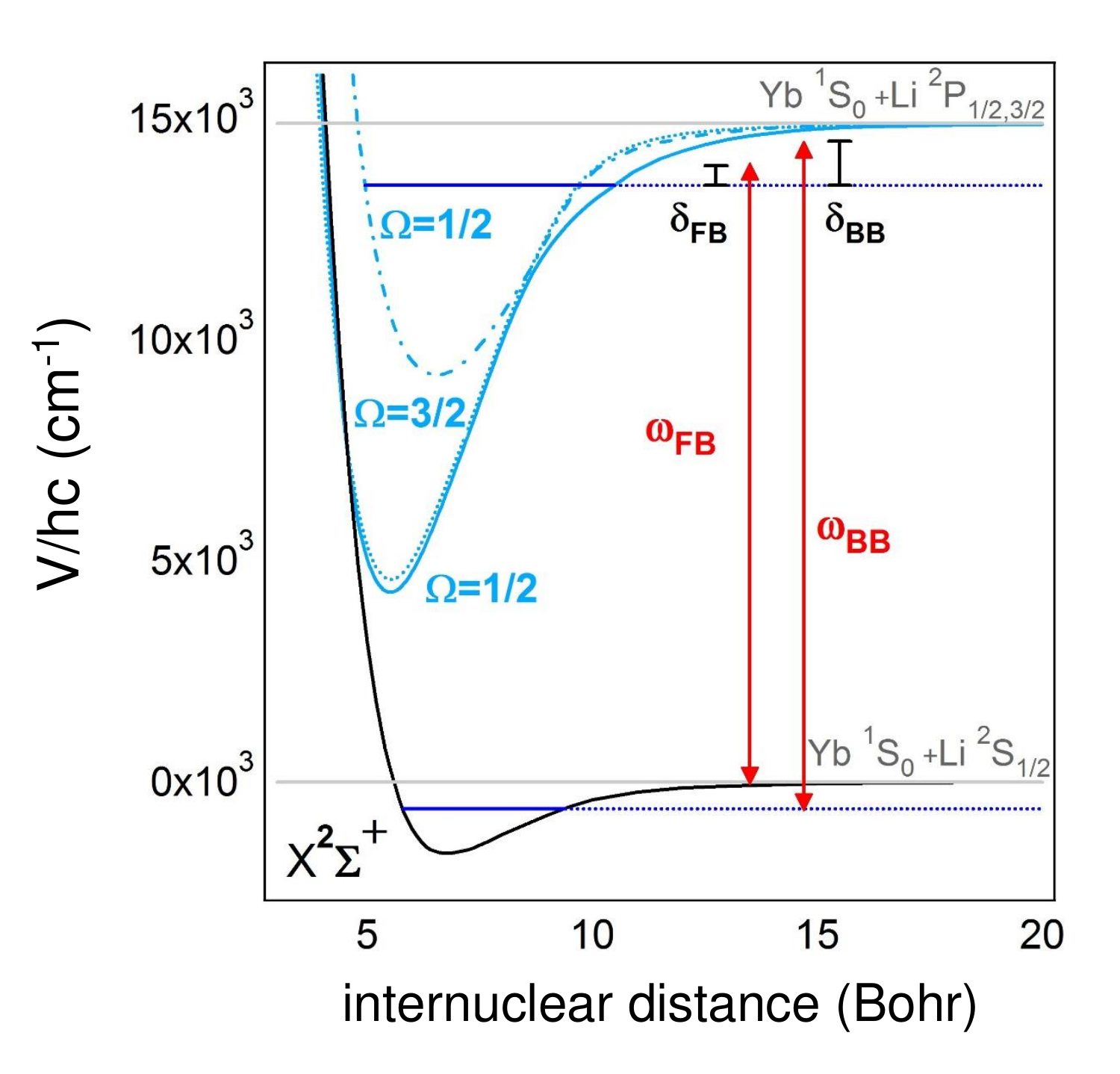}
\par\end{centering}
\caption{A plot of the relevant interatomic potentials with a schematic representation of two-photon spectroscopy. The potentials are plotted to scale while the locations of the bound states are exaggerated for clarity. $\omega_{FB}$ and $\omega_{BB}$ are the angular frequencies of the free-bound and bound-bound lasers respectively. $\delta_{FB}$ and $\delta_{BB}$ are the one-photon detunings from the free-bound and bound-bound PA resonances respectively.}
\label{fig:schematic}
\end{figure}

Two-photon PA spectroscopy can be used to probe the interatomic potential of an initially unbound atom pair~\cite{abra95,tsai97,dutt17,gutt18}. The schematic for the PA process is shown in Figure~\ref{fig:schematic}. Within a trapped mixture of ultracold atoms, a Li and Yb atom pair is resonantly coupled to an electronically excited molecular state by a free-bound (FB) laser. A bound-bound (BB) laser couples this electronically excited molecule to the electronic ground state. Resonant two-photon coupling is achieved when the energy difference between the BB and FB photons is equal to the binding energy of the ground-state molecule.

We detect resonant coupling to a molecular state by trap-loss spectroscopy. When the atoms are exposed to only the FB laser beam, a one-photon free-bound PA resonance is detected as a frequency-dependent loss of atoms from the trap. For two-photon spectroscopy, we stabilize the frequency of the FB laser to a free-bound PA resonance, thus inducing a fixed atom loss from the trap via molecule formation for a fixed exposure time. We monitor this loss in the additional presence of the BB laser beam. As the BB laser frequency is varied, a bound-bound resonance is detected as a suppression of this loss, caused by strong coupling between the excited and ground molecular states which shifts the energy of the excited molecular state and takes the FB laser off the free-bound resonance~\cite{jone06}. 

In this work we interrogate the least bound states of the YbLi $^{2}\Sigma^+$ molecular ground potential via an excited molecular state within the Yb($^{1}S_0$) + Li($^{2}P$) electronic manifold. The intermediate state belongs to a combination of two $\Omega=1/2$ and an $\Omega=3/2$ relativisitic potentials which asymptotically correspond to the Yb($^{1}S_0$) + Li($^{2}P_{1/2}$) and the Yb($^{1}S_0$) + Li($^{2}P_{3/2}$) thresholds~\cite{roy16PA}. $\Omega$ is the total projection quantum number onto the interatomic axis. Guided by observed favorable free-bound Franck-Condon factors, we induce trap loss through one-photon PA by specifically addressing either the second or the seventh least bound states (labeled $v^*=-2,-7$) within a single vibrational series in the intermediate state~\footnote{We will present our spectroscopy of this excited molecular state, which extends our previous observations in~\cite{roy16PA}, in a future publication.}. The bound-bound Franck-Condon factors for either of these two intermediate states were high enough to observe the six least-bound vibrational states (labeled $v$=-1 to -6) in the $^{2}\Sigma^+$ electronic ground state.

\begin{figure*}
\centering
\captionsetup{position=top, justification=raggedright, singlelinecheck=false, labelfont=large}
    \subfloat[]{\includegraphics[width=0.32\textwidth]{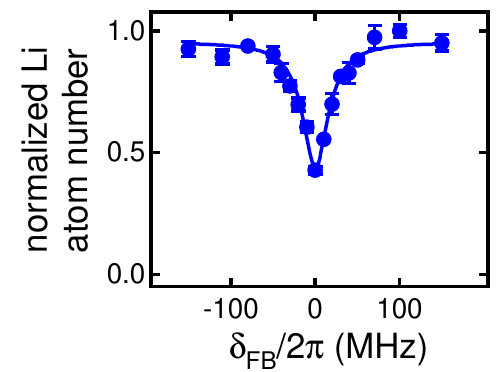}}
    \hfill
    \subfloat[\label{subfig:lifetime}]{\includegraphics[width=0.32\textwidth]{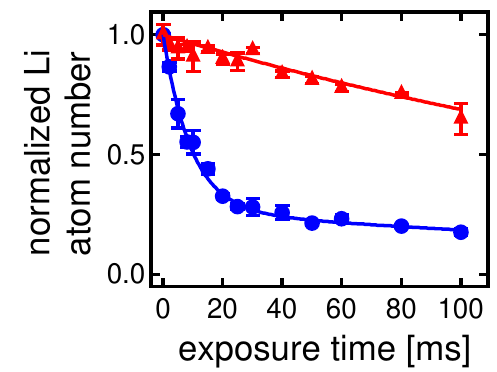}}
    \hfill
    \subfloat[\label{subfig:twophoton}]{\includegraphics[width=0.32\textwidth]{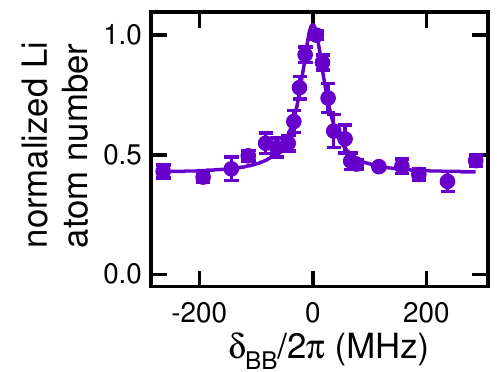}}
    \caption{
    \label{figure:sampledata}
    (a) A representative one-photon PA loss resonance of $v^*=-2$ with a binding energy of 15.9 GHz relative to the Yb $^{1}\text{S}_0 +$Li $^{2}\text{P}_{\nicefrac{1}{2}}$ free atom energy. The solid line is a Lorentzian fit. (b) Evolution of the Li atom number in the presence of resonant (blue circles) and off-resonant (red triangles) PA light. In the resonant case the FB laser is locked to the $v^*=-2$. In the off-resonant case it is locked $2 \pi \times$1 GHz to the red of resonance. The off-resonant data is fit with an exponential and the on-resonant data is fit with a biexponential for which one of the time constants is fixed to that of the fit to the off-resonant data. (c) A representative two-photon PA resonance of $v=-3$ with binding energy 123.96 GHz relative to the Yb $^{1}\text{S}_0 +$Li $^{2}\text{S}_{\nicefrac{1}{2}}$ free atom energy. The solid line is a Lorentzian fit. All error bars are statistical.}
\end{figure*}

\subsection{Experimental setup}

Our $^{174}$Yb-$^6$Li mixture is prepared through a sequence of laser and evaporative cooling techniques using procedures similar to those described in~\cite{roy17}. For the measurements reported in this paper, the atoms are confined in a crossed optical dipole trap with mean harmonic frequency ${\bar \omega}_{\rm Yb(Li)} = 2 \pi \times 492 (3936)\,$Hz, and with tightest confinement along the vertical direction. Prior to application of PA light, the mixture population of $N_{\text{Yb}}=2.0 \times 10^6$ in the $^1S_0$ ground state and $N_{\text{Li}}=1.4 \times 10^5$ in the $^2S_{1/2}$ $\ket{m_J=-\frac{1}{2},m_I=-1}$ ground state, has a common temperature of $T=15\,\mu$K. The clouds are above quantum degeneracy with $T/T_{\rm c} \simeq 5 $ and $T/T_{\rm F} \simeq 1$, where $T_{\rm c}$ is the Yb BEC transition temperature and $T_{\rm F}$ is the Li Fermi temperature. The peak atomic density is $n_{\rm Yb(Li)} = 2 \times 10^{14} (3.9 \times 10^{13})\,$cm$^{-3}$.

The PA light is derived from two diode master lasers, each of which is amplified by a tapered amplifier (TA). Using a scanning Fabry-Perot cavity, the frequency of the free-bound laser is actively stabilized relative to an additional laser, itself stabilized to an atomic reference. A wavemeter (High Finesse WS-7) is used to stabilize the frequency of the bound-bound laser. The power in each beam is controlled using independent acousto-optic modulators (AOMs). The two beams are combined and passed through a heated Li vapor cell to suppress frequency components near the Li atomic resonances generated by the TA. The spectrally filtered beams are coupled into a single-mode polarization-maintaining fiber, the output of which is focused onto the trapped atomic mixture with a waist of 75 $\mu \text{m}$. This combined beam provided up to 170 $\nicefrac{\text{W}}{\text{cm}^2}$ in the FB and 280 $\nicefrac{\text{W}}{\text{cm}^2}$ in the BB but the intensities were adjusted for optimum signal, accommodating different Franck-Condon factors. After an exposure time ranging from 10 to 100 ms, the number of atoms remaining in the trap is measured using absorption imaging. A typical 1-photon PA resonance is shown in Fig.~\ref{figure:sampledata}(a).

The spectroscopic search for heteronuclear YbLi resonances can be hindered by background losses from the trap, notably homonuclear photoassociation of $\text{Li}_2$~\cite{roy16PA}. We exploit the fermionic nature of $^{6}\text{Li}$ to mitigate this background loss through the use of a spin polarized sample at a temperature well below the $p$-wave barrier for the Li-Li scattering system ($\sim 8\,$mK). We perform our spectroscopic search using $\sigma^-$ PA beam polarizations at a bias field of $420$ G. Off-resonant scattering from the PA lasers thus addresses a $m_I+m_J=-\frac{3}{2} \longrightarrow -\frac{5}{2}$ cycling transition, preserving the initial spin polarization. The viability of two-photon spectroscopy with such a setup is demonstrated in Fig.~\ref{figure:sampledata}(b). The loss of atoms when the FB laser is off resonance is sufficiently small compared to the resonant loss to allow for a strong two-photon trap-loss signal. The off-resonant loss is dominated by spontaneous scattering despite nearby $\text{Li}_2$ resonances centered at $2 \pi \times$240, 320 and -220 MHz relative to the $v^*=-2$ YbLi resonance~\cite{abra95PA}.

The small reduced mass of YbLi leads to a large vibrational spacing of molecular levels, posing another challenge for the spectroscopic search. Detecting the least-bound vibrational state $v=-1$ required a thorough search over several GHz of spectral range. We utilized the Leroy-Bernstein relation~\cite{lero70} as well as an earlier calculation of the C$_6$ coefficient for this potential~\cite{Roy2016}  to predict the approximate binding energy of each subsequent state, thus expediting our search. We used this approach to detect weakly bound vibrational states with binding energies extending to greater than 1 THz.

\subsection{Long Range Bound States}

A typical two-photon resonance is shown in Figure~\ref{figure:sampledata}(c). The measured binding energies of the six rovibrational states observed in this way are presented in Table~\ref{tb:GrndBnd} \footnote{Deeper bound states ($v=-7$ and lower) were inaccessible in this work due to the choice of laser diodes used for PA. Future work can expand on the current study by using laser diodes centered at longer wavelengths.}. The associated error is a combination of PA laser linewidth and wavemeter precision, except for the shallowest state, where a higher precision is achieved through coherent two-photon spectroscopy and is discussed in Section~\ref{sec:darkstatespec}. 

\begin{figure}
\begin{centering}
\includegraphics[width=1\columnwidth]{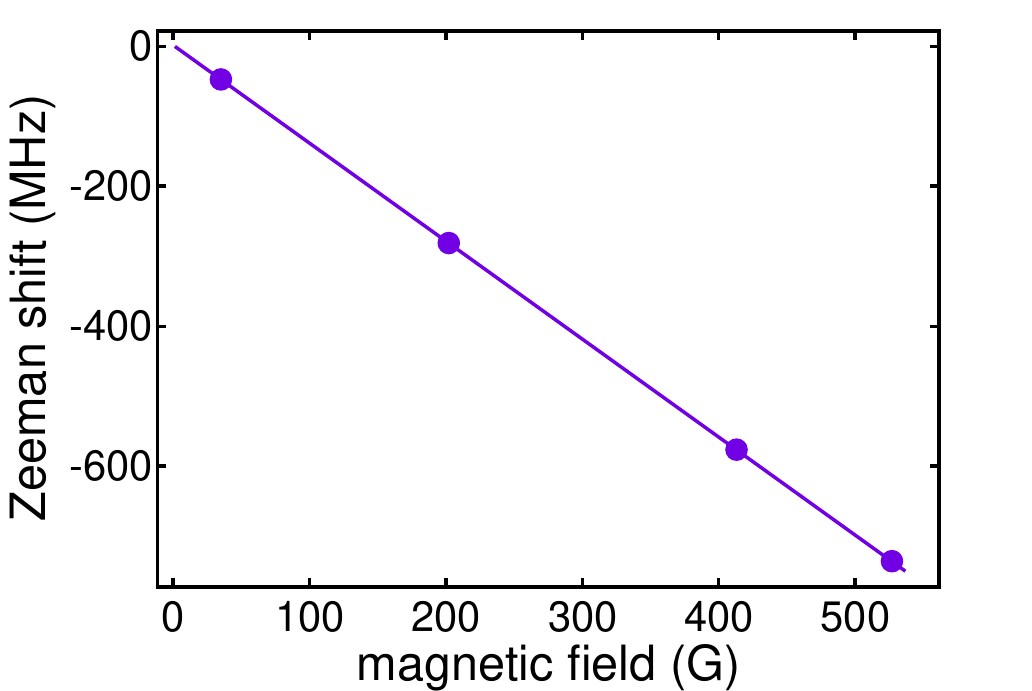}
\par\end{centering}
\caption{ 
\label{fig:magmoment}
The magnetic moment of the shallowest bound state of $X^2\Sigma^+$ for $^{174}$Yb$^{6}$Li is determined by measuring its binding energy at various magnetic fields using two-photon PA spectroscopy. At each field, the well-known Zeeman shift of the atomic state is subtracted from the apparent shift in the binding energy to give the Zeeman shift of the molecule.}
\end{figure}

As discussed in Section~\ref{sec:theory}, the observed spectrum corresponds to a single vibrational series from $v=-1$ to $v=-6$. In order to verify the magnetic quantum number of these states, we determine the magnetic moment of the $v=-1$ state using two-photon PA spectroscopy at various magnetic fields (see Fig.~\ref{fig:magmoment}). The magnetic moment is measured to be 1.40(16) MHz/G, equal to that of the initial atom pair state.

To determine the rotational state of the observed levels, we first note that since the experimental temperature is well below the {\it p}-wave barrier of $\sim 3$ mK for the Yb-Li scattering system, we can exclude {\it p}-wave and any higher partial wave collisions. This leaves the atoms to interact dominantly through the $l=0$ $s$-wave channel. Using the conservation of total angular momentum and parity along with the fact that the excited intermediate state has an electronic orbital angular momentum $L=1$, we can conclude that the final rovibrational state after the two-photon PA can only be $l=0$ or $2$. Further identification of the rotational quantum number is realized by fitting the binding energies to a theoretical model, detailed in Section~\ref{sec:theory}, which determines that all observed bound states have $l=0$ \footnote{The frequency range for our search included the expected frequency of the $v=-1$, $l=2$ resonance but our signal to noise ratio was not sufficient to observe it, likely because of low Franck-Condon factors.}.

\section{\label{sec:theory}Theoretical Analysis: Ground Potential and State Assignments}

\subsection{Ground electronic potential}

\begin{figure}
 \includegraphics[width=0.9\columnwidth,trim=0 20 10 60,clip]{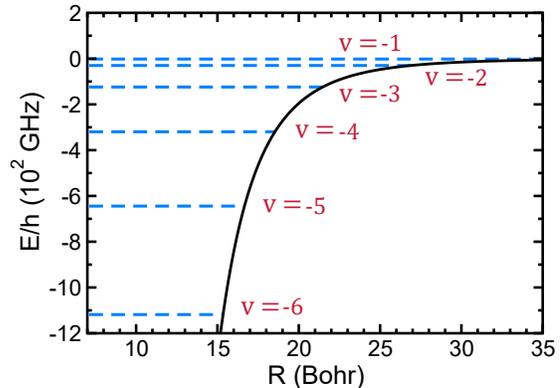}
 \caption{The long-range shape of the $X^2\Sigma^{+}$ potential of YbLi
 and the measured weakly bound rovibrational states labeled by horizontal
 dashed lines.}
 \label{fig:GrndPot}
\end{figure}

Ground state Li$(^{2}$S$_{1/2})$ and Yb$(^{1}$S$_0)$ atoms interact through the X$^2\Sigma^{+}$ Born-Oppenheimer molecular potential. Its weakly bound rovibrational states are particularly sensitive to the long-range part of the potential. We use the Tang-Toennies damped long-range form~\cite{Tang84}
\begin{equation}
V(R)=-\sum_{n=3}^5 f_{2n}(R)\frac{C_{2n}}{R^{2n}} \;,
\label{eq:VLR}
\end{equation}
where $R$ is the internuclear separation, $f_{2n}(R)$ is a damping function of the form 
\begin{equation}
f_{2n}(R) = 1-\left(\sum_{k=0}^{2n}
            \frac{(bR)^k}{k!}\right)e^{-bR} \;,
\end{equation}
and $b$ is a fitting parameter that is related to the ionization energies of the two atoms. The leading term of Eq.~(\ref{eq:VLR}) is defined by the $C_6$ van der Waals coefficient that has been calculated in our previous work to be $1496(150) E_ha_0^6$~\cite{Roy2016}. It agrees with other theoretical predictions within our uncertainty, {\it i.e.} $1594 E_ha_0^6$ from Ref.~\cite{Zhang2010} and~\cite{Brue2013}, and $1551(31) E_ha_0^6$ from Ref.~\cite{Safronova2012}. Here $E_h$ is the Hartree energy and $a_0$ is the Bohr radius. 
The next term is described by the $C_8$ coefficient which was estimated by Porsev {\it et al.} to be $1.27(3) \times 10^5 E_ha_0^8$~\cite{Porsev2014}. We use $C_{10}=(49/40)\times C_8^2/C_6$ recommended by
Thakkar and Smith~\cite{Thakkar1974} to reduce the number of fitting parameters. After adjusting the coefficients to fit the measured energies of the weakly bound states, we obtain $C_6=1581(5) E_ha_0^6$, $C_8=1.49(2)\times 10^5 E_ha_0^8$, and $b=0.912(9) a_0^{-1}$ (see Section~\ref{sec:theoryb}). The coefficients are given in atomic units which are abbreviated to a.u. in the following text. Figure~\ref{fig:GrndPot} shows the fitted long-range potential and the measured energies of the weakly bound states. The precise shape of the short-range part of the potential plays a less significant role in determining the weakly bound state energies. We use a short-range potential that agrees with the average well depth and the general shape of the {\it ab initio} potentials from~\cite{Gopakumar2010,Zhang2010,Kotochigova2011,Tohme2015}, and smoothly connect it to the long-range part at $14$ $a_0$. 

The depth of the short-range well is adjusted with a Gaussian function
\begin{equation}
g(R) = V_d\times e^{-\frac{(R-R_0)^2}{2c^2}} \;
\end{equation}
to facilitate the fitting of the measured energies, where $c^2=1.5$ $a_0^2$ is the minimum of the potential well and $V_d$ is the adjustable parameter for the depth of the potential well at the minimum. When adjusting $V_d$, the total number of vibrational states $N_v$ is kept the same ($N_v=24$ for the current potential). It is possible to determine $N_v$ through a sufficiently precise measurement of the least bound state isotope shift. However such a measurement is outside the scope of this work.

\subsection{\label{sec:theoryb}Weakly bound rovibrational states}

To calculate the energy of a weakly bound state in the ground electronic potential, we use the spinless basis set $|L,m_L\rangle Y_{lm}(\hat{R})$ to expand its wavefunction, where $L=0$ is the electronic orbital angular momentum with $m_L=0$ its projection in the laboratory frame, and $Y_{lm}(\hat{R})$ is the spherical harmonic of relative orbital angular momentum $l$ and its projection $m$. The unit vector $\hat{R}$ is the direction of the molecular axis in the laboratory frame. The molecular Hamiltonian is then diagonal in this basis set and is the same for different $m$
\footnote{
We neglect small effects such as couplings between the rotational angular momentum of the diatom and the spins, including the nuclear and the electronic ones, and the dependence on the nuclear separation $R$ of the hyperfine constant and the {\it g}-factors. Thus the spins are completely decoupled from the nuclear motion, and the hyperfine splittings of the molecular bound states in a magnetic field is the same as the ones in the separate-atoms limit.}. We choose $m=0$ for all our calculations. We use a discrete variable representation (DVR) based method to solve the single channel radial Schr\"{o}dinger equation numerically for the rovibrational bound state energies of $v=-1$ to $-6$. Since the binding energies measured experimentally are the only levels observed for each vibrational quantum number, we cannot directly assign rotational quantum numbers to them. Thus, we vary
$l$ from $0$ to $2$ to serve as an unknown variable in our fitting procedure.

\begin{table}
\caption{\label{tb:GrndBnd} Comparisons between the experimentally observed bound state energies and the theory results of $l=0$, 1, and 2. $h$ is the Planck constant and the frequencies are given in GHz. The theoretical uncertainties in the last digit are given in parenthesis which are derived from error propagation using the covariance matrix of the four parameters of the potential.}
\begin{ruledtabular}
\begin{tabular}{ccccc}
$v$ & exp. $E/h$ & $E^{th}_{l=0}/h$ & 
  $E^{th}_{l=1}/h$ & $E^{th}_{l=2}/h$ \\\hline
$-1$ &   -1.8260(15) &   -1.8260(34) &   -1.2933(34) &   -0.3214(33) \\
$-2$ &   -30.070(26) &   -30.100(17) &   -28.740(16) &   -26.047(14) \\
$-3$ &  -123.957(26) &  -124.002(43) &  -121.840(40) &  -117.533(36) \\
$-4$ &  -319.465(26) &  -319.406(45) &  -316.470(43) &  -310.610(42) \\
$-5$ &  -644.457(26) &  -644.484(57) &  -640.812(60) &  -633.477(64) \\
$-6$ & -1118.295(26) & -1118.285(60) & -1113.927(59) & -1105.219(58) \\
$-7$ &       -       &  -1754.07(51) &  -1749.08(51) &  -1739.10(49) \\
$-8$ &       -       &   -2546.7(23) &   -2541.1(23) &   -2530.0(23) \\
\end{tabular}
\end{ruledtabular}
\end{table}

To assign rotational quantum number $l_v$ to the observed states $v$ as well as to find the best fit for the four parameters of the potential including the two dispersion coefficients, the damping function fitting parameter $b$, and the depth of the short-range potential well $V_d$, we adopt the following procedure. First, we choose a set of $C_6$, $C_8$, $b$, and $l_{v=-1}$, and vary $V_d$ within a range where the total number of bound states remains the same. $V_d$ is varied until the binding energy of the chosen $l_{v=-1}$ state approaches the observed $v=-1$ energy within $1\times 10^{-5}$ GHz. Next, we calculate the bound state energies of $v=-2$ to $-6$ and $l_v=0$, $1$, and $2$, and for each $v$ we choose among the set of $l_v$ the one closest to the observed energy. For each $v$, we record the chosen $l_v$ and the energy difference $\Delta E_v$ from the observed value. Then we can define $\chi^2$ as
\begin{equation}
\chi^2 = \sum_{v=1}^6 
    \frac{\Delta E_{v}^2}{u_v^2} \;,
\end{equation}
where $u_v$ is the experimental uncertainty associated with the measured binding energy $E_{v}$ of the vibrational state $v$. We also define a total rotation number
\begin{equation}
\lambda = \sum_{v=-1}^{-6} l_{v}\times 3^{v+6} \;
\label{eq:totrot}
\end{equation}
so that every different set of $\{l_{v}\}$ corresponds to a unique $\lambda$ value. In particular, $\lambda=0$ when all six states have $l_{v}=0$.

\begin{figure*}
\centering
\captionsetup{position=top, justification=raggedright, singlelinecheck=false, labelfont=large}
    \subfloat[\label{fig:fitgrndpt}]{\includegraphics[width=0.49\textwidth]{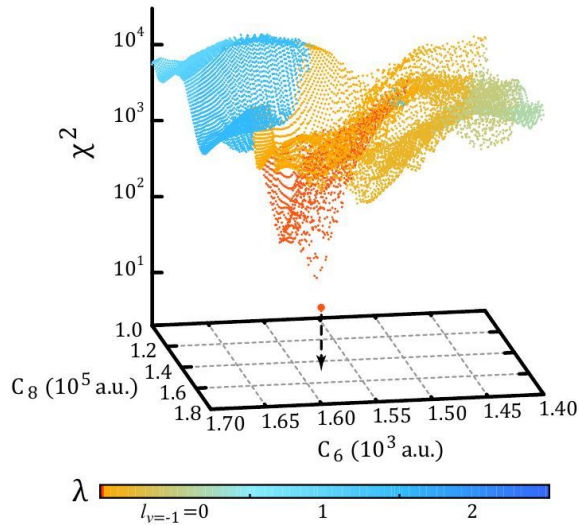}}
    \hfill
    \subfloat[\label{fig:fitgrnddt}]{\includegraphics[width=0.49\textwidth]{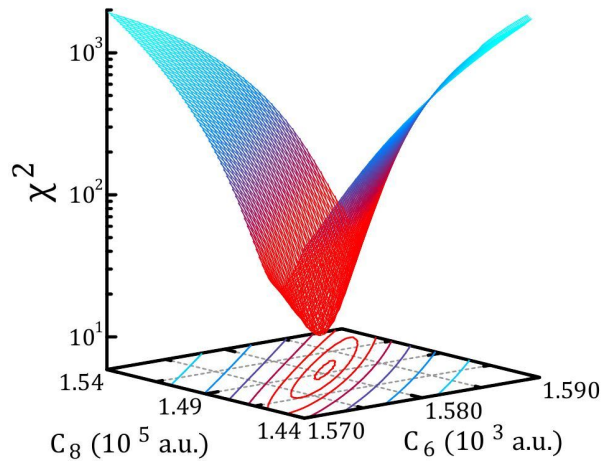}}
    \caption{(color online) $\chi^2$ as a function of the $C_6$ and $C_8$ parameters. The rotational quantum number $l$, the $b$ parameter, and the depth of the potential well are all used to minimize  $\chi^2$ for a specific set of $C_6$ and  $C_8$. Panel (a) shows the initial search with a coarse grid. The color is coded to the total rotation number $\lambda$ defined in Eq.~(\ref{eq:totrot}) according to the color bar. The left side of the color bar corresponds to smaller $\lambda$ and the right side corresponds to larger $\lambda$, with $\lambda=0$ specifically coded in deep orange. According to Eq.~(\ref{eq:totrot}), we divide the color bar of $\lambda$ into three sections. The one on the left corresponds to $l_{v=-1}=0$, the one in the middle corresponds to $l_{v=-1}=1$, and the one on the right corresponds to $l_{v=-1}=2$. The point with the lowest $\chi^2$ is magnified and the dashed arrow points to its location on the 2D grid of $C_6$ and $C_8$. Panel (b) shows the second step of the search with a much finer grid in the adjacent area of the minimum of the initial search. The three dimensional surface of $\chi^2$ as well as its contours are color coded with the value of $\chi^2$. Red indicates smaller values while light blue indicates larger values.}
    \label{fig:fitgrnd}
\end{figure*}

To minimize $\chi^2$ in a large multidimensional parameter space, we opt to search with coarse step sizes first, then in the vicinity of the minimum we conduct a second search with much finer step sizes. In the first step, we search the $C_6$ coefficient from $1400$ a.u. to $2100$ a.u. with a step size of $2$ a.u., the $C_8$ coefficient from $1\times 10^5$ a.u. to $1.8\times 10^5$ a.u. with a step size of $1\times 10^3$ a.u., and the $b$ parameter from $0.7$ a.u. to $1.2$ a.u. with a step size of $0.005$ a.u. $\chi^2$ calculated from the first step is plotted against $C_6$ and $C_8$ coefficients in Figure~\ref{fig:fitgrnd}a. Parameter $b$ is chosen to minimize $\chi^2$ for each pair of $C_6$ and $C_8$. The color is coded proportional to $\lambda$ according to the color bar in Figure~\ref{fig:fitgrnd}a with $\lambda = 0$ in dark orange. Data points with $\chi^2$ lower than $\sim 10^2$ all corresponds to $\lambda = 0$. The lowest $\chi^2$ is $12.3$ that corresponds to $C_6=1580$ a.u.,
$C_8=1.49\times 10^5$ a.u., and $b=0.910$ a.u. The result of the first step shows that all the 
experimentally observed bound states correspond to $l=0$ rotational state.

In the second step, we search the vicinity of the point with minimal $\chi^2$ in the first step with finer step sizes. 
We vary the $C_6$ coefficient from $1560$ a.u. to $1600$ a.u. with a step size of $0.2$ a.u., the $C_8$ coefficient from $1.40\times 10^5$ a.u. to $1.57\times 10^5$ a.u. with a step size of $50$ a.u., and the $b$ parameter from $0.88$ a.u. to $0.93$ a.u. with a step size of $2\times 10^{-4}$ a.u. $\chi^2$ calculated from the second step is plotted in Figure~\ref{fig:fitgrnd}b. Again, parameter $b$ is chosen to minimize $\chi^2$ for each pair of $C_6$ and $C_8$. The colors of the surface as well as the contours on the 2D grid are coded with the value of $\chi^2$ with lower values approaching red and higher values approaching light blue. All the points shown in the graph corresponds to $\lambda=0$ which confirms our initial assignment of $l=0$ for all states. The lowest $\chi^2$ is $10.69$ with $C_6=1581.0$ a.u., $C_8=1.4945\times 10^5$ a.u., and $b=0.9116$ a.u. Table~\ref{tb:GrndBnd} shows the comparisons between the observed energy levels and the calculated ones with the given parameters. The energies predicted for $l=1$ and $2$ as well as $v=-7$ and $-8$ are also presented.

To analyze the uncertainties of and the correlations between the fitted parameters, we follow the error propagation analysis of Ref.~\cite{CODATA1998}. We have six experimental measurements and four continuous fitting variables~\footnote{Although $l_v$ are used as unknown variables, they are discrete and remain zero in the vicinity of the minimum where the error propagation is practically done. Therefore we don't count them as variables in the error analysis.}. Thus we have two degrees of freedom left. The lowest $\chi^2$ of $10.69$ then corresponds to a reduced $\chi^2$ of $5.35$ and a Birge ratio $R_\mathrm{B}$ of $2.31$. Since our Birge ratio is larger than one, we uniformly inflate the experimental uncertainty by the Birge ratio ($u_v\times R_\mathrm{B}$). Following the methodology in the Appendix E and F of Ref.~\cite{CODATA1998}, we arrive at the uncertainties of and correlation coefficients between the four derived parameters $C_6$, $C_8$, $b$, and $V_d$ shown in Table~\ref{tb:cov}. The four parameters are highly correlated, indicated by the fact that all the correlation coefficients are close to one. 

\section{\label{sec:scatlength} ground state s-wave scattering length}

Using the fitted potential, we can calculate the zero energy $s$-wave scattering length $a_{s0}$ numerically. At a finite energy, the $s$-wave scattering length in single channel scattering is defined as
\begin{equation}
a_s(E)=\frac{1}{ik}\times\frac{1-S}{1+S} \;,
\end{equation}
where $k=\sqrt{2\mu E}/\hbar$ is the wavenumber with $\mu$ the reduced mass and $\hbar=h/2\pi$ the reduced Planck constant, and $S$ is the complex single channel scattering matrix. The zero energy $s$-wave scattering length is then $a_{s0}=\lim_{E\to 0}a_s(E)$. We numerically propagate the single channel radial Schr\"{o}dinger equation using Gordon's propagator~\cite{Gordon69}, and match the wavefunction at large $R$ to the scattering boundary condition to obtain the $S$ matrix. As a result, the zero energy $s$-wave scattering length is $30.074(55)$ $a_0$. The uncertainty in the parenthesis is propagated from the variances and covariances of the four potential parameters. 

\begin{table}
\caption{\label{tb:cov} The uncertainties of and correlation coefficients between the values of the fitted parameters. The uncertainties are the diagonal elements in bold and are in atomic units apart from the one for $V_d$, which is in cm$^{-1}$. The dimensionless 
correlation coefficients are symmetric and defined as $r(x_i,x_j)=u(x_i,x_j)/[u(x_i)u(x_j)]$, where $u(x_i,x_j)$ is the covariance between parameters $x_i$ and $x_j$ and 
$u(x_{i,j})$ are the uncertainties of parameters $x_i$ and $x_j$ respectively. Only the upper-triangle of the correlation matrix is shown.}
\begin{ruledtabular}
\begin{tabular}{ccccc}
         &  $C_6$ &  $C_8$ & $b$ & $V_d$ \\\hline
   $C_6$ & $\bm{5.18}$ & 0.920 & 0.993 & 0.973\\
   $C_8$ &   & $\bm{1.91\times 10^3}$ & 0.944 & 0.984\\
     $b$ &   &   & $\bm{8.81\times 10^{-3}}$ & 0.987 \\
   $V_d$ &   &   &   & $\bm{11.52}$ \\
\end{tabular}
\end{ruledtabular}
\end{table}

\begin{figure}
 \includegraphics[width=0.9\columnwidth,trim=0 0 0 0,clip]{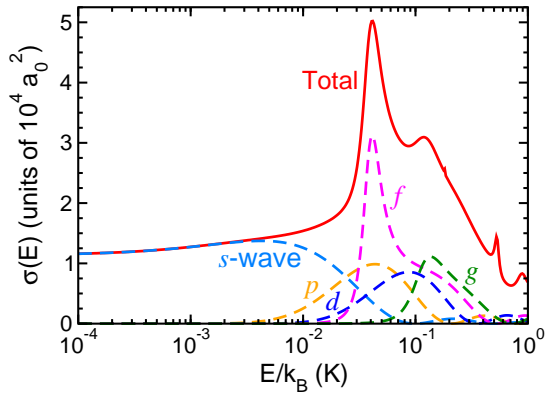}
 \caption{(color online) The total and first five partial elastic scattering cross sections as functions of collision energy. The solid red line corresponds to the total cross section. The dashed lines correspond to partial cross sections with $l=0$ to 4, which are labelled as $s$- to $g$-wave. Note that not all partial wave cross sections that contribute significantly to the total cross section are shown here.}
 \label{fig:GrndSc}
\end{figure}

We also calculate the total elastic cross sections $\sigma_\mathrm{tot}(E)=\Sigma_l\sigma_l(E)$, where
\begin{equation}
\sigma_l(E) = \frac{\pi}{k^2}(2l+1)|1-S|^2
\end{equation}
are the partial cross sections and the summation over $l$ is from 0 to infinity. The $s$-wave partial cross section is related to the $s$-wave scattering length, $\sigma_{l=0}(E)=4\pi a_s(E)^2$. Figure~\ref{fig:GrndSc} shows the total cross section as well as the first five partial cross sections as functions of collision energy. The large peak near $0.04$ K comes from a shape resonance in the $l=3$ channel. 

The $a_{s0}$ obtained here is $\sim 50\%$ larger than the previous results determined from interspecies thermalization experiments~\cite{Ivanov2011,Hara2011}. The extraction of scattering length from thermalization measurements is sensitive to systematic effects stemming from the determination of absolute densities and interspecies spatial overlap. While photoassociation spectroscopy together with detailed theoretical analysis as presented in this paper provide more precise values, to gain further confidence in this updated value of the scattering length, we have performed some additional checks as described below.  

\subsection{Quantum defect theory predictions}

The quantum defect theory (QDT) for $-C_\alpha/R^\alpha$ type of long-range potentials with $\alpha>2$~\cite{Gao2008} provides an analytic connection between the energies of the weakly bound states and the low energy scattering properties through the short-range parameters that are slowly varying functions of energy near the threshold.  In particular, the least bound state energy is related to $a_{s0}$ through the short-range $K$-matrix, $K_c(E_s,l)$~\cite{Gao2008}. $E_s=E/s_E$ is the dimensionless scaled energy with the energy scale $s_E=(\hbar^2/2\mu)(1/\beta_6)^2$, where the length scale $\beta_6=(2\mu C_6/\hbar^2)^{1/4}$. For $^6$Li$^{174}$Yb with $C_6=1581.0 E_h/a_0^6$, $\beta_6=76.1$ $a_0$ and $s_E/h=53.6$ MHz. These two scales are relatively 
insensitive to the change of $C_6$. The {\it s}-wave scattering length is related to the zero energy zero partial wave $K_c$ by
\begin{equation}
a_{s0}=\beta_6\left(b^{2b}\frac{\Gamma(1-b)}{\Gamma(1+b)}\right)
    \frac{K_c(0,0)+\tan(\pi b/2)}{K_c(0,0)-\tan(\pi b/2)} \;,
    \label{eq:as0}
\end{equation}
where $b=1/(\alpha-2)$ with $\alpha=6$. The least bound state energy is the negative root of the equation
\begin{equation}
\chi_l^c(E_s)=K_c(l,E_s)
\label{eq:chic}
\end{equation}
that is the closest to zero. $\chi_l^c$ is a QDT special function defined in Ref.~\cite{Gao2008}. At $E=-1.8260$ GHz, the least bound state energy corresponds to the scaled energy $E_s=-34.1$. Since $K_c$ varies slowly with energy~\cite{Gao2008}, we can approximate $K_c(0,0)$ with $K_c(E_s,0)$. Thus, using Eqs.~(\ref{eq:chic}) and
(\ref{eq:as0}), we get $a_{0s}=29.2$ $a_0$, which is consistent with our numerical result above. The slight discrepancy comes from the small energy dependence of the $K_c$ parameter that is mainly due 
to the effect of the $-C_8/R^8$ term that modifies the long-range wavefunction between the length scale of the $-C_8/R^8$ term and the length scale of the $-C_6/R^6$ term, {\it i.e.} $\beta_6$. 
The QDT result confirms that the connection between $a_{s0}$ and the least bound energy is independent of the detailed shape of the short-range potential and the total number of vibrational states it supports.

\section{\label{sec:darkstatespec}Coherent Dark State Spectroscopy}

In this section, we present our results on the observation of atom-molecule coherence in the YbLi system through coherent two-photon spectroscopy of the least-bound vibrational state $v=-1$. In this scheme, free atom pairs are coupled to this electronic ground molecular state via an excited molecular state with a pair of phase-coherent lasers. This realizes a coherent three-level system where it is well-known that a dark superposition state can be formed~\cite{harr97}. In the context of PA, the dark state is evident in the suppression of PA loss in a frequency range much narrower than the one-photon PA transition itself~\cite{wink05}. The location of this narrow line provides a more precise measurement of the $v=-1$ binding energy. Additionally, our work demonstrates the existence of ground state YbLi molecules in the mixture, and is an important step towards efficient molecule formation using StiRAP from free atoms~\cite{stel12}. 

To achieve the necessary phase stability between the optical drive fields involved in the coherent two-photon spectroscopy, we modify the experimental setup discussed in Sec~\ref{sec:twophspec} in the following way. We derive the FB and the BB laser beams by splitting the output of a single master diode laser and generating the required frequency difference using AOMs. The $\sim 1.8\,$GHz binding energy of the $v=-1$ state is bridged using the BB (FB) AOM in an upshifted (downshifted) 4-pass orientation~\cite{luwa17} and driving both at a frequency $\sim 225\,$MHz using the same direct-digital synthesis (DDS) radiofrequency source.

\begin{figure}
\begin{centering}
\includegraphics[width=1\columnwidth]{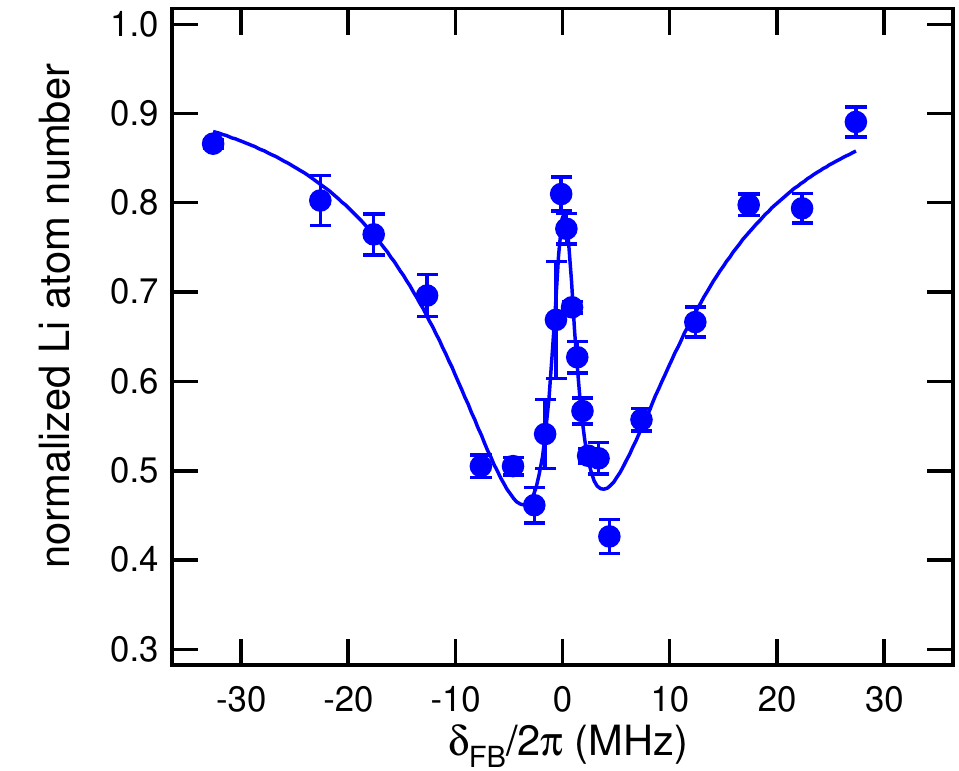}
\par\end{centering}
\caption{ 
\label{fig:darkstate}
Representative coherent two-photon spectrum in which the ground $v=-1$ state is observed as the narrow atom-molecule dark-state feature within a broad loss resonance. For this data, the BB laser was fixed at the $v*=-2 \leftrightarrow v=-1$ resonance while the FB laser was varied across the $v*=-2$ 1-photon PA resonance. The solid line is a sum of two Lorentzians fit to the data. The error bars are statistical.}
\end{figure}

A representative dark-state spectrum is shown in Figure~\ref{fig:darkstate}, where a sub-natural linewidth of $2\,$MHz is observed using the $v*=-2$ intermediate state. By reducing the BB power and using $v*=-7$ intermediate state, we have observed linewidths below $400\,$kHz with reduced strength of the dark resonance feature. Possible causes for the observed linewidth include atom-molecule collisions, spontaneous scattering, and finite temperature effects. The co-existing molecular fraction could be improved in a 3D optical lattice with doublon site filling, where efficient coherent free-bound StiRAP into ground state YbLi molecules may be possible, as has been demonstrated in the homonuclear Sr$_2$ system~\cite{stel12,ciam17}.

\section{\label{sec:summary} Summary and Outlook}

We have reported our results on two-photon photoassociation spectroscopy of the weakly bound vibrational levels in the electronic ground state of the $^{174}$Yb$^6$Li molecule. Combining our experimental findings with a detailed theoretical study, we obtained accurate values for the long-range dispersion coefficients for the inter-atomic potential. 

Our work also yields information on Yb-Li scattering properties. We obtained an updated value for the Yb-Li $s$-wave scattering length of $30a_0$. Furthermore, these results can refine the predicted locations of magnetic Feshbach resonances in the Yb-Li collisional system~\cite{brue12} \footnote{We have recently observed magnetic Feshbach resonances in Yb-Li at locations consistent with predictions based on the work presented in this paper. These observations will be reported in a forthcoming publication.}. Such resonances characteristic of ground state scattering between alkali and alkaline-earth-like atoms~\cite{barb18} could be used for molecule formation~\cite{mark18}.

The observation of coherent atom-molecule dark states in YbLi is an important step towards efficient coherent molecule formation through all-optical methods, possibly in 3D optical lattice confinement~\cite{stel12,ciam17}. These results together with advances in the related systems of RbYb, RbSr, CsYb~\cite{munc11,ciam18,gutt18,barb18}, provide promise for the realization of a new class of ultracold molecules in the $^2\Sigma^+$ ground state, beyond the bi-alkali paradigm, with applications in quantum simulation and information processing, precision measurements and ultracold chemistry.

\section{Acknowledgments} 
We thank Ryan Bowler for helpful discussion and experimental contribution in the initial stages of this work. We further thank Xinxin Tang and Khang Ton for experimental assistance. Research at University of Washington is supported by NSF Grant No. PHY-1806212 and AFOSR Grant No.FA9550-19-1-0012. Work at Temple University is supported by the ARO Grant No. W911NF-17-1-0563, the AFOSR Grant No. FA9550-14-1-0321, and the NSF Grant No. PHY-1619788.


\begin{thebibliography}{55}%
\makeatletter
\providecommand \@ifxundefined [1]{%
 \@ifx{#1\undefined}
}%
\providecommand \@ifnum [1]{%
 \ifnum #1\expandafter \@firstoftwo
 \else \expandafter \@secondoftwo
 \fi
}%
\providecommand \@ifx [1]{%
 \ifx #1\expandafter \@firstoftwo
 \else \expandafter \@secondoftwo
 \fi
}%
\providecommand \natexlab [1]{#1}%
\providecommand \enquote  [1]{``#1''}%
\providecommand \bibnamefont  [1]{#1}%
\providecommand \bibfnamefont [1]{#1}%
\providecommand \citenamefont [1]{#1}%
\providecommand \href@noop [0]{\@secondoftwo}%
\providecommand \href [0]{\begingroup \@sanitize@url \@href}%
\providecommand \@href[1]{\@@startlink{#1}\@@href}%
\providecommand \@@href[1]{\endgroup#1\@@endlink}%
\providecommand \@sanitize@url [0]{\catcode `\\12\catcode `\$12\catcode
  `\&12\catcode `\#12\catcode `\^12\catcode `\_12\catcode `\%12\relax}%
\providecommand \@@startlink[1]{}%
\providecommand \@@endlink[0]{}%
\providecommand \url  [0]{\begingroup\@sanitize@url \@url }%
\providecommand \@url [1]{\endgroup\@href {#1}{\urlprefix }}%
\providecommand \urlprefix  [0]{URL }%
\providecommand \Eprint [0]{\href }%
\providecommand \doibase [0]{https://doi.org/}%
\providecommand \selectlanguage [0]{\@gobble}%
\providecommand \bibinfo  [0]{\@secondoftwo}%
\providecommand \bibfield  [0]{\@secondoftwo}%
\providecommand \translation [1]{[#1]}%
\providecommand \BibitemOpen [0]{}%
\providecommand \bibitemStop [0]{}%
\providecommand \bibitemNoStop [0]{.\EOS\space}%
\providecommand \EOS [0]{\spacefactor3000\relax}%
\providecommand \BibitemShut  [1]{\csname bibitem#1\endcsname}%
\let\auto@bib@innerbib\@empty
\bibitem [{\citenamefont {Carr}\ \emph {et~al.}(2009)\citenamefont {Carr},
  \citenamefont {DeMille}, \citenamefont {Krems},\ and\ \citenamefont
  {Ye}}]{carr09}%
  \BibitemOpen
  \bibfield  {author} {\bibinfo {author} {\bibfnamefont {L.~D.}\ \bibnamefont
  {Carr}}, \bibinfo {author} {\bibfnamefont {D.}~\bibnamefont {DeMille}},
  \bibinfo {author} {\bibfnamefont {R.~V.}\ \bibnamefont {Krems}},\ and\
  \bibinfo {author} {\bibfnamefont {J.}~\bibnamefont {Ye}},\ }\bibfield
  {title} {\bibinfo {title} {Cold and ultracold molecules: science, technology,
  and applications},\ }\href@noop {} {\bibfield  {journal} {\bibinfo  {journal}
  {New J. Phys.}\ }\textbf {\bibinfo {volume} {11}},\ \bibinfo {pages} {055049}
  (\bibinfo {year} {2009})}\BibitemShut {NoStop}%
\bibitem [{\citenamefont {Gadway}\ and\ \citenamefont {Yan}(2016)}]{gadw16}%
  \BibitemOpen
  \bibfield  {author} {\bibinfo {author} {\bibfnamefont {B.}~\bibnamefont
  {Gadway}}\ and\ \bibinfo {author} {\bibfnamefont {B.}~\bibnamefont {Yan}},\
  }\bibfield  {title} {\bibinfo {title} {Strongly interacting ultracold polar
  molecules},\ }\href@noop {} {\bibfield  {journal} {\bibinfo  {journal} {J.
  Phys. B.}\ }\textbf {\bibinfo {volume} {49}},\ \bibinfo {pages} {152002}
  (\bibinfo {year} {2016})}\BibitemShut {NoStop}%
\bibitem [{\citenamefont {Zelevinsky}\ \emph {et~al.}(2008)\citenamefont
  {Zelevinsky}, \citenamefont {Kotochigova},\ and\ \citenamefont
  {Ye}}]{zele08}%
  \BibitemOpen
  \bibfield  {author} {\bibinfo {author} {\bibfnamefont {T.}~\bibnamefont
  {Zelevinsky}}, \bibinfo {author} {\bibfnamefont {S.}~\bibnamefont
  {Kotochigova}},\ and\ \bibinfo {author} {\bibfnamefont {J.}~\bibnamefont
  {Ye}},\ }\bibfield  {title} {\bibinfo {title} {Precision test of mass-ratio
  variations with lattice-confined ultracold molecules},\ }\href@noop {}
  {\bibfield  {journal} {\bibinfo  {journal} {Phys. Rev. Lett.}\ }\textbf
  {\bibinfo {volume} {100}},\ \bibinfo {pages} {043201} (\bibinfo {year}
  {2008})}\BibitemShut {NoStop}%
\bibitem [{\citenamefont {Hudson}\ \emph {et~al.}(2008)\citenamefont {Hudson},
  \citenamefont {Gilfoy}, \citenamefont {Kotochigova}, \citenamefont {Sage},\
  and\ \citenamefont {DeMille}}]{huds08}%
  \BibitemOpen
  \bibfield  {author} {\bibinfo {author} {\bibfnamefont {E.~R.}\ \bibnamefont
  {Hudson}}, \bibinfo {author} {\bibfnamefont {N.~B.}\ \bibnamefont {Gilfoy}},
  \bibinfo {author} {\bibfnamefont {S.}~\bibnamefont {Kotochigova}}, \bibinfo
  {author} {\bibfnamefont {J.~M.}\ \bibnamefont {Sage}},\ and\ \bibinfo
  {author} {\bibfnamefont {D.}~\bibnamefont {DeMille}},\ }\bibfield  {title}
  {\bibinfo {title} {Inelastic collisions of ultracold heteronuclear molecules
  in an optical trap},\ }\href@noop {} {\bibfield  {journal} {\bibinfo
  {journal} {Phys. Rev. Lett.}\ }\textbf {\bibinfo {volume} {100}},\ \bibinfo
  {pages} {203201} (\bibinfo {year} {2008})}\BibitemShut {NoStop}%
\bibitem [{\citenamefont {Kajita}(2008)}]{kaji08}%
  \BibitemOpen
  \bibfield  {author} {\bibinfo {author} {\bibfnamefont {M.}~\bibnamefont
  {Kajita}},\ }\bibfield  {title} {\bibinfo {title} {Prospects of detecting $\nicefrac{m_e}{m_p}$} variance using vibrational transition frequencies of $^{2}\Sigma-$state molecules,\ }\href
  {https://doi.org/10.1103/PhysRevA.77.012511} {\bibfield  {journal} {\bibinfo
  {journal} {Phys. Rev. A}\ }\textbf {\bibinfo {volume} {77}},\ \bibinfo
  {pages} {012511} (\bibinfo {year} {2008})}\BibitemShut {NoStop}%
\bibitem [{\citenamefont {Flambaum}\ \emph {et~al.}(2014)\citenamefont
  {Flambaum}, \citenamefont {DeMille},\ and\ \citenamefont {Kozlov}}]{flaum14}%
  \BibitemOpen
  \bibfield  {author} {\bibinfo {author} {\bibfnamefont {V.~V.}\ \bibnamefont
  {Flambaum}}, \bibinfo {author} {\bibfnamefont {D.}~\bibnamefont {DeMille}},\
  and\ \bibinfo {author} {\bibfnamefont {M.~G.}\ \bibnamefont {Kozlov}},\
  }\bibfield  {title} {\bibinfo {title} {Time-reversal symmetry violation in
  molecules induced by nuclear magnetic quadrupole moments},\ }\href
  {https://doi.org/10.1103/PhysRevLett.113.103003} {\bibfield  {journal}
  {\bibinfo  {journal} {Phys. Rev. Lett.}\ }\textbf {\bibinfo {volume} {113}},\
  \bibinfo {pages} {103003} (\bibinfo {year} {2014})}\BibitemShut {NoStop}%
\bibitem [{\citenamefont {Hudson}\ \emph {et~al.}(2011)\citenamefont {Hudson},
  \citenamefont {Kara}, \citenamefont {Smallman}, \citenamefont {Sauer},
  \citenamefont {Tarbutt},\ and\ \citenamefont {Hinds}}]{huds11}%
  \BibitemOpen
  \bibfield  {author} {\bibinfo {author} {\bibfnamefont {J.}~\bibnamefont
  {Hudson}}, \bibinfo {author} {\bibfnamefont {D.}~\bibnamefont {Kara}},
  \bibinfo {author} {\bibfnamefont {I.}~\bibnamefont {Smallman}}, \bibinfo
  {author} {\bibfnamefont {B.}~\bibnamefont {Sauer}}, \bibinfo {author}
  {\bibfnamefont {M.}~\bibnamefont {Tarbutt}},\ and\ \bibinfo {author}
  {\bibfnamefont {E.}~\bibnamefont {Hinds}},\ }\bibfield  {title} {\bibinfo
  {title} {Improved measurement of the shape of the electron},\ }\href
  {https://doi.org/https://doi.org/10.1038/nature10104} {\bibfield  {journal}
  {\bibinfo  {journal} {Nature}\ }\textbf {\bibinfo {volume} {473}},\ \bibinfo
  {pages} {493} (\bibinfo {year} {2011})}\BibitemShut {NoStop}%
\bibitem [{\citenamefont {Anderegg}\ \emph {et~al.}(2018)\citenamefont
  {Anderegg}, \citenamefont {Augenbraun}, \citenamefont {Bao}, \citenamefont
  {Burchesky}, \citenamefont {Cheuk}, \citenamefont {Ketterle},\ and\
  \citenamefont {Doyle}}]{ande18}%
  \BibitemOpen
  \bibfield  {author} {\bibinfo {author} {\bibfnamefont {L.}~\bibnamefont
  {Anderegg}}, \bibinfo {author} {\bibfnamefont {B.~L.}\ \bibnamefont
  {Augenbraun}}, \bibinfo {author} {\bibfnamefont {Y.}~\bibnamefont {Bao}},
  \bibinfo {author} {\bibfnamefont {S.}~\bibnamefont {Burchesky}}, \bibinfo
  {author} {\bibfnamefont {L.}~\bibnamefont {Cheuk}}, \bibinfo {author}
  {\bibfnamefont {W.}~\bibnamefont {Ketterle}},\ and\ \bibinfo {author}
  {\bibfnamefont {J.}~\bibnamefont {Doyle}},\ }\bibfield  {title} {\bibinfo
  {title} {Laser cooling of optically trapped molecules},\ }\href@noop {}
  {\bibfield  {journal} {\bibinfo  {journal} {Nat. Phys.}\ }\textbf {\bibinfo
  {volume} {14}},\ \bibinfo {pages} {890} (\bibinfo {year}
  {2018})}\BibitemShut {NoStop}%
\bibitem [{\citenamefont {Ni}\ \emph {et~al.}(2008)\citenamefont {Ni},
  \citenamefont {Ospelkaus}, \citenamefont {de~Miranda}, \citenamefont {Peer},
  \citenamefont {Neyenhuis}, \citenamefont {Zirbel}, \citenamefont
  {Kotochigova}, \citenamefont {Julienne}, \citenamefont {Jin},\ and\
  \citenamefont {Ye}}]{ni08}%
  \BibitemOpen
  \bibfield  {author} {\bibinfo {author} {\bibfnamefont {K.-K.}\ \bibnamefont
  {Ni}}, \bibinfo {author} {\bibfnamefont {S.}~\bibnamefont {Ospelkaus}},
  \bibinfo {author} {\bibfnamefont {M.~H.~G.}\ \bibnamefont {de~Miranda}},
  \bibinfo {author} {\bibfnamefont {A.}~\bibnamefont {Peer}}, \bibinfo {author}
  {\bibfnamefont {B.}~\bibnamefont {Neyenhuis}}, \bibinfo {author}
  {\bibfnamefont {J.~J.}\ \bibnamefont {Zirbel}}, \bibinfo {author}
  {\bibfnamefont {S.}~\bibnamefont {Kotochigova}}, \bibinfo {author}
  {\bibfnamefont {P.~S.}\ \bibnamefont {Julienne}}, \bibinfo {author}
  {\bibfnamefont {D.~S.}\ \bibnamefont {Jin}},\ and\ \bibinfo {author}
  {\bibfnamefont {J.}~\bibnamefont {Ye}},\ }\bibfield  {title} {\bibinfo
  {title} {A high phase-space-density gas of polar molecules},\ }\href@noop {}
  {\bibfield  {journal} {\bibinfo  {journal} {Science}\ }\textbf {\bibinfo
  {volume} {322}},\ \bibinfo {pages} {231} (\bibinfo {year}
  {2008})}\BibitemShut {NoStop}%
\bibitem [{\citenamefont {Takekoshi}\ \emph {et~al.}(2014)\citenamefont
  {Takekoshi}, \citenamefont {Reichs$\ddot{\rm o}$llner}, \citenamefont
  {Schindewolf}, \citenamefont {Hutson}, \citenamefont {Le~Sueur},
  \citenamefont {Dulieu}, \citenamefont {Ferlaino}, \citenamefont {Grimm},\
  and\ \citenamefont {N$\ddot{\rm a}$gerl}}]{take14}%
  \BibitemOpen
  \bibfield  {author} {\bibinfo {author} {\bibfnamefont {T.}~\bibnamefont
  {Takekoshi}}, \bibinfo {author} {\bibfnamefont {L.}~\bibnamefont
  {Reichs$\ddot{\rm o}$llner}}, \bibinfo {author} {\bibfnamefont
  {A.}~\bibnamefont {Schindewolf}}, \bibinfo {author} {\bibfnamefont
  {J.~M.}~\bibnamefont {Hutson}}, \bibinfo {author} {\bibfnamefont
  {C.~R.}~\bibnamefont {Le~Sueur}}, \bibinfo {author} {\bibfnamefont
  {O.}~\bibnamefont {Dulieu}}, \bibinfo {author} {\bibfnamefont
  {F.}~\bibnamefont {Ferlaino}}, \bibinfo {author} {\bibfnamefont
  {R.}~\bibnamefont {Grimm}},\ and\ \bibinfo {author} {\bibfnamefont {H.-C.}\
  \bibnamefont {N$\ddot{\rm a}$gerl}},\ }\bibfield  {title} {\bibinfo {title}
  {Ultracold dense samples of dipolar RbCs molecules in the rovibrational and
  hyperfine ground state},\ }\href@noop {} {\bibfield  {journal} {\bibinfo
  {journal} {Phys. Rev. Lett.}\ }\textbf {\bibinfo {volume} {113}},\ \bibinfo
  {pages} {205301} (\bibinfo {year} {2014})}\BibitemShut {NoStop}%
\bibitem [{\citenamefont {Molony}\ \emph {et~al.}(2014)\citenamefont {Molony},
  \citenamefont {Gregory}, \citenamefont {Ji}, \citenamefont {Lu},
  \citenamefont {K$\ddot{\rm o}$ppinger}, \citenamefont {Le~Sueur},
  \citenamefont {Blackley}, \citenamefont {Hutson},\ and\ \citenamefont
  {Cornish}}]{molo14}%
  \BibitemOpen
  \bibfield  {author} {\bibinfo {author} {\bibfnamefont {P.~K.}~\bibnamefont
  {Molony}}, \bibinfo {author} {\bibfnamefont {P.~D.}~\bibnamefont {Gregory}},
  \bibinfo {author} {\bibfnamefont {Z.}~\bibnamefont {Ji}}, \bibinfo {author}
  {\bibfnamefont {B.}~\bibnamefont {Lu}}, \bibinfo {author} {\bibfnamefont
  {M.~P.}~\bibnamefont {K$\ddot{\rm o}$ppinger}}, \bibinfo {author} {\bibfnamefont
  {C.~R.}~\bibnamefont {Le~Sueur}}, \bibinfo {author} {\bibfnamefont
  {C.~L.}~\bibnamefont {Blackley}}, \bibinfo {author} {\bibfnamefont
  {J.~M.}~\bibnamefont {Hutson}},\ and\ \bibinfo {author} {\bibfnamefont
  {S.~L.}~\bibnamefont {Cornish}},\ }\bibfield  {title} {\bibinfo {title}
  {Creation of ultracold $^{87}$Rb$^{133}$Cs molecules in the rovibrational ground
  state},\ }\href@noop {} {\bibfield  {journal} {\bibinfo  {journal} {Phys.
  Rev. Lett.}\ }\textbf {\bibinfo {volume} {113}},\ \bibinfo {pages} {255301}
  (\bibinfo {year} {2014})}\BibitemShut {NoStop}%
\bibitem [{\citenamefont {Park}\ \emph {et~al.}(2015)\citenamefont {Park},
  \citenamefont {Will},\ and\ \citenamefont {Zwierlein}}]{park15}%
  \BibitemOpen
  \bibfield  {author} {\bibinfo {author} {\bibfnamefont {J.~W.}~\bibnamefont
  {Park}}, \bibinfo {author} {\bibfnamefont {S.~A.}~\bibnamefont {Will}},\ and\
  \bibinfo {author} {\bibfnamefont {M.~W.}~\bibnamefont {Zwierlein}},\ }\bibfield
  {title} {\bibinfo {title} {Ultracold dipolar gas of fermionic $^{23}$Na$^{40}$K
  molecules in their absolute ground state},\ }\href@noop {} {\bibfield
  {journal} {\bibinfo  {journal} {Phys. Rev. Lett.}\ }\textbf {\bibinfo
  {volume} {114}},\ \bibinfo {pages} {205302} (\bibinfo {year}
  {2015})}\BibitemShut {NoStop}%
\bibitem [{\citenamefont {Guo}\ \emph {et~al.}(2016)\citenamefont {Guo},
  \citenamefont {Zhu}, \citenamefont {Lu}, \citenamefont {Ye}, \citenamefont
  {Wang}, \citenamefont {Vexiau}, \citenamefont {Bouloufa-Maafa}, \citenamefont
  {Qu\'em\'ener}, \citenamefont {Dulieu},\ and\ \citenamefont {Wang}}]{Guo16}%
  \BibitemOpen
  \bibfield  {author} {\bibinfo {author} {\bibfnamefont {M.}~\bibnamefont
  {Guo}}, \bibinfo {author} {\bibfnamefont {B.}~\bibnamefont {Zhu}}, \bibinfo
  {author} {\bibfnamefont {B.}~\bibnamefont {Lu}}, \bibinfo {author}
  {\bibfnamefont {X.}~\bibnamefont {Ye}}, \bibinfo {author} {\bibfnamefont
  {F.}~\bibnamefont {Wang}}, \bibinfo {author} {\bibfnamefont {R.}~\bibnamefont
  {Vexiau}}, \bibinfo {author} {\bibfnamefont {N.}~\bibnamefont
  {Bouloufa-Maafa}}, \bibinfo {author} {\bibfnamefont {G.}~\bibnamefont
  {Qu\'em\'ener}}, \bibinfo {author} {\bibfnamefont {O.}~\bibnamefont
  {Dulieu}},\ and\ \bibinfo {author} {\bibfnamefont {D.}~\bibnamefont {Wang}},\
  }\bibfield  {title} {\bibinfo {title} {Creation of an ultracold gas of
  ground-state dipolar $^{23}\mathrm{Na}^{87}\mathrm{Rb}$ molecules},\ }\href
  {https://doi.org/10.1103/PhysRevLett.116.205303} {\bibfield  {journal}
  {\bibinfo  {journal} {Phys. Rev. Lett.}\ }\textbf {\bibinfo {volume} {116}},\
  \bibinfo {pages} {205303} (\bibinfo {year} {2016})}\BibitemShut {NoStop}%
\bibitem [{\citenamefont {Rvachov}\ \emph {et~al.}(2017)\citenamefont
  {Rvachov}, \citenamefont {Son}, \citenamefont {Sommer}, \citenamefont
  {Ebadi}, \citenamefont {Park}, \citenamefont {Zwierlein}, \citenamefont
  {Ketterle},\ and\ \citenamefont {Jamison}}]{rvac17}%
  \BibitemOpen
  \bibfield  {author} {\bibinfo {author} {\bibfnamefont {T.~M.}\ \bibnamefont
  {Rvachov}}, \bibinfo {author} {\bibfnamefont {H.}~\bibnamefont {Son}},
  \bibinfo {author} {\bibfnamefont {A.~T.}\ \bibnamefont {Sommer}}, \bibinfo
  {author} {\bibfnamefont {S.}~\bibnamefont {Ebadi}}, \bibinfo {author}
  {\bibfnamefont {J.~J.}\ \bibnamefont {Park}}, \bibinfo {author}
  {\bibfnamefont {M.~W.}\ \bibnamefont {Zwierlein}}, \bibinfo {author}
  {\bibfnamefont {W.}~\bibnamefont {Ketterle}},\ and\ \bibinfo {author}
  {\bibfnamefont {A.~O.}\ \bibnamefont {Jamison}},\ }\bibfield  {title}
  {\bibinfo {title} {Long-lived ultracold molecules with electric and magnetic
  dipole moments},\ }\href {https://doi.org/10.1103/PhysRevLett.119.143001}
  {\bibfield  {journal} {\bibinfo  {journal} {Phys. Rev. Lett.}\ }\textbf
  {\bibinfo {volume} {119}},\ \bibinfo {pages} {143001} (\bibinfo {year}
  {2017})}\BibitemShut {NoStop}%
\bibitem [{\citenamefont {Brue}\ and\ \citenamefont {Hutson}(2012)}]{brue12}%
  \BibitemOpen
  \bibfield  {author} {\bibinfo {author} {\bibfnamefont {D.~A.}~\bibnamefont
  {Brue}}\ and\ \bibinfo {author} {\bibfnamefont {J.~M.}~\bibnamefont {Hutson}},\
  }\bibfield  {title} {\bibinfo {title} {Magnetically tunable Feshbach
  resonances in ultracold Li-Yb mixtures},\ }\href@noop {} {\bibfield
  {journal} {\bibinfo  {journal} {Phys. Rev. Lett.}\ }\textbf {\bibinfo
  {volume} {108}},\ \bibinfo {pages} {043201} (\bibinfo {year}
  {2012})}\BibitemShut {NoStop}%
\bibitem [{\citenamefont {Dowd}\ \emph {et~al.}(2015)\citenamefont {Dowd},
  \citenamefont {Roy}, \citenamefont {Shrestha}, \citenamefont {Petrov},
  \citenamefont {Makrides}, \citenamefont {Kotochigova},\ and\ \citenamefont
  {Gupta}}]{dowd15}%
  \BibitemOpen
  \bibfield  {author} {\bibinfo {author} {\bibfnamefont {W.}~\bibnamefont
  {Dowd}}, \bibinfo {author} {\bibfnamefont {R.}~\bibnamefont {Roy}}, \bibinfo
  {author} {\bibfnamefont {R.}~\bibnamefont {Shrestha}}, \bibinfo {author}
  {\bibfnamefont {A.}~\bibnamefont {Petrov}}, \bibinfo {author} {\bibfnamefont
  {C.}~\bibnamefont {Makrides}}, \bibinfo {author} {\bibfnamefont
  {S.}~\bibnamefont {Kotochigova}},\ and\ \bibinfo {author} {\bibfnamefont
  {S.}~\bibnamefont {Gupta}},\ }\bibfield  {title} {\bibinfo {title} {Magnetic
  field dependent interactions in an ultracold Li-Yb(${^3P}_2$) mixture},\
  }\href@noop {} {\bibfield  {journal} {\bibinfo  {journal} {New J. Phys.}\
  }\textbf {\bibinfo {volume} {17}},\ \bibinfo {pages} {055007} (\bibinfo
  {year} {2015})}\BibitemShut {NoStop}%
\bibitem [{\citenamefont {Barb\'e}\ \emph {et~al.}(2018)\citenamefont {Barb\'e},
  \citenamefont {Ciamei}, \citenamefont {Pasquiou}, \citenamefont
  {Reichs$\ddot{\rm o}$llner}, \citenamefont {Schreck}, \citenamefont {$\dot{\rm Z}$uchowski},\ and\
  \citenamefont {Hutson}}]{barb18}%
  \BibitemOpen
  \bibfield  {author} {\bibinfo {author} {\bibfnamefont {V.}~\bibnamefont
  {Barb\'e}}, \bibinfo {author} {\bibfnamefont {A.}~\bibnamefont {Ciamei}},
  \bibinfo {author} {\bibfnamefont {B.}~\bibnamefont {Pasquiou}}, \bibinfo
  {author} {\bibfnamefont {L.}~\bibnamefont {Reichs$\ddot{\rm o}$llner}}, \bibinfo {author}
  {\bibfnamefont {F.}~\bibnamefont {Schreck}}, \bibinfo {author} {\bibfnamefont
  {P.}~\bibnamefont {$\dot{\rm Z}$uchowski}},\ and\ \bibinfo {author} {\bibfnamefont
  {J.~M.}\ \bibnamefont {Hutson}},\ }\bibfield  {title} {\bibinfo {title}
  {Observation of Feshbach resonances between alkali and closed-shell atoms},\
  }\href {https://doi.org/https://doi.org/10.1038/s41567-018-0169-x} {\bibfield
   {journal} {\bibinfo  {journal} {Nature Physics}\ }\textbf {\bibinfo {volume}
  {14}},\ \bibinfo {pages} {881} (\bibinfo {year} {2018})}\BibitemShut
  {NoStop}%
\bibitem [{\citenamefont {M$\ddot{\rm u}$nchow}\ \emph
  {et~al.}(2011)\citenamefont {M$\ddot{\rm u}$nchow}, \citenamefont {Bruni},
  \citenamefont {Madalinski},\ and\ \citenamefont {G$\ddot{\rm
  o}$rlitz}}]{munc11}%
  \BibitemOpen
  \bibfield  {author} {\bibinfo {author} {\bibfnamefont {F.}~\bibnamefont
  {M$\ddot{\rm u}$nchow}}, \bibinfo {author} {\bibfnamefont {C.}~\bibnamefont
  {Bruni}}, \bibinfo {author} {\bibfnamefont {M.}~\bibnamefont {Madalinski}},\
  and\ \bibinfo {author} {\bibfnamefont {A.}~\bibnamefont {G$\ddot{\rm
  o}$rlitz}},\ }\bibfield  {title} {\bibinfo {title} {Two-photon spectroscopy
  of heteronuclear YbRb},\ }\href@noop {} {\bibfield  {journal} {\bibinfo
  {journal} {Phys. Chem. Chem. Phys.}\ }\textbf {\bibinfo {volume} {13}},\
  \bibinfo {pages} {18734} (\bibinfo {year} {2011})}\BibitemShut {NoStop}%
\bibitem [{\citenamefont {Roy}\ \emph {et~al.}(2016{\natexlab{a}})\citenamefont
  {Roy}, \citenamefont {Shrestha}, \citenamefont {Green}, \citenamefont
  {Gupta}, \citenamefont {Li}, \citenamefont {Kotochigova}, \citenamefont
  {Petrov},\ and\ \citenamefont {Yuen}}]{roy16PA}%
  \BibitemOpen
  \bibfield  {author} {\bibinfo {author} {\bibfnamefont {R.}~\bibnamefont
  {Roy}}, \bibinfo {author} {\bibfnamefont {R.}~\bibnamefont {Shrestha}},
  \bibinfo {author} {\bibfnamefont {A.}~\bibnamefont {Green}}, \bibinfo
  {author} {\bibfnamefont {S.}~\bibnamefont {Gupta}}, \bibinfo {author}
  {\bibfnamefont {M.}~\bibnamefont {Li}}, \bibinfo {author} {\bibfnamefont
  {S.}~\bibnamefont {Kotochigova}}, \bibinfo {author} {\bibfnamefont
  {A.}~\bibnamefont {Petrov}},\ and\ \bibinfo {author} {\bibfnamefont {C.~H.}\
  \bibnamefont {Yuen}},\ }\bibfield  {title} {\bibinfo {title}
  {Photoassociative production of ultracold heteronuclear ${\mathrm{YbLi}}^{*}$
  molecules},\ }\href {https://doi.org/10.1103/PhysRevA.94.033413} {\bibfield
  {journal} {\bibinfo  {journal} {Phys. Rev. A}\ }\textbf {\bibinfo {volume}
  {94}},\ \bibinfo {pages} {033413} (\bibinfo {year}
  {2016}{\natexlab{a}})}\BibitemShut {NoStop}%
\bibitem [{\citenamefont {{Ciamei}}\ \emph {et~al.}(2018)\citenamefont
  {{Ciamei}}, \citenamefont {{Szczepkowski}}, \citenamefont {{Bayerle}},
  \citenamefont {{Barb\'e}}, \citenamefont {{Reichs$\ddot{\rm o}$llner}}, \citenamefont
  {{Tzanova}}, \citenamefont {{Chen}}, \citenamefont {{Pasquiou}},
  \citenamefont {{Grochola}}, \citenamefont {{Kowalczyk}}, \citenamefont
  {{Jastrzebski}},\ and\ \citenamefont {{Schreck}}}]{ciam18}%
  \BibitemOpen
  \bibfield  {author} {\bibinfo {author} {\bibfnamefont {A.}~\bibnamefont
  {{Ciamei}}}, \bibinfo {author} {\bibfnamefont {J.}~\bibnamefont
  {{Szczepkowski}}}, \bibinfo {author} {\bibfnamefont {A.}~\bibnamefont
  {{Bayerle}}}, \bibinfo {author} {\bibfnamefont {V.}~\bibnamefont {{Barb\'e}}},
  \bibinfo {author} {\bibfnamefont {L.}~\bibnamefont {{Reichs$\ddot{\rm o}$llner}}},
  \bibinfo {author} {\bibfnamefont {S.~M.}\ \bibnamefont {{Tzanova}}}, \bibinfo
  {author} {\bibfnamefont {C.-C.}\ \bibnamefont {{Chen}}}, \bibinfo {author}
  {\bibfnamefont {B.}~\bibnamefont {{Pasquiou}}}, \bibinfo {author}
  {\bibfnamefont {A.}~\bibnamefont {{Grochola}}}, \bibinfo {author}
  {\bibfnamefont {P.}~\bibnamefont {{Kowalczyk}}}, \bibinfo {author}
  {\bibfnamefont {W.}~\bibnamefont {{Jastrzebski}}},\ and\ \bibinfo {author}
  {\bibfnamefont {F.}~\bibnamefont {{Schreck}}},\ }\bibfield  {title} {\bibinfo
  {title} {{The RbSr $^{2}\Sigma^{+}$ ground state investigated via spectroscopy of hot and ultracold molecules}},\ }\href@noop {} {\bibfield  {journal} {\bibinfo
  {journal} {Phys. Chem. Chem. Phys.}\ }\textbf {\bibinfo {volume} {20}},\
  \bibinfo {pages} {26221} (\bibinfo {year} {2018})}\BibitemShut {NoStop}%
\bibitem [{\citenamefont {Guttridge}\ \emph {et~al.}(2018)\citenamefont
  {Guttridge}, \citenamefont {Frye}, \citenamefont {Yang}, \citenamefont
  {Hutson},\ and\ \citenamefont {Cornish}}]{gutt18}%
  \BibitemOpen
  \bibfield  {author} {\bibinfo {author} {\bibfnamefont {A.}~\bibnamefont
  {Guttridge}}, \bibinfo {author} {\bibfnamefont {M.~D.}\ \bibnamefont {Frye}},
  \bibinfo {author} {\bibfnamefont {B.~C.}\ \bibnamefont {Yang}}, \bibinfo
  {author} {\bibfnamefont {J.~M.}\ \bibnamefont {Hutson}},\ and\ \bibinfo
  {author} {\bibfnamefont {S.~L.}\ \bibnamefont {Cornish}},\ }\bibfield
  {title} {\bibinfo {title} {Two-photon photoassociation spectroscopy of CsYb:
  Ground-state interaction potential and interspecies scattering lengths},\
  }\href@noop {} {\bibfield  {journal} {\bibinfo  {journal} {Phys. Rev. A}\
  }\textbf {\bibinfo {volume} {98}},\ \bibinfo {pages} {022707} (\bibinfo
  {year} {2018})}\BibitemShut {NoStop}%
\bibitem [{\citenamefont {Micheli}\ \emph {et~al.}(2006)\citenamefont
  {Micheli}, \citenamefont {Brennen},\ and\ \citenamefont {Zoller}}]{mich06}%
  \BibitemOpen
  \bibfield  {author} {\bibinfo {author} {\bibfnamefont {A.}~\bibnamefont
  {Micheli}}, \bibinfo {author} {\bibfnamefont {G.~K.}\ \bibnamefont
  {Brennen}},\ and\ \bibinfo {author} {\bibfnamefont {P.}~\bibnamefont
  {Zoller}},\ }\bibfield  {title} {\bibinfo {title} {A toolbox for lattice-spin
  models with polar molecules},\ }\href@noop {} {\bibfield  {journal} {\bibinfo
   {journal} {Nature Physics}\ }\textbf {\bibinfo {volume} {2}},\ \bibinfo
  {pages} {341} (\bibinfo {year} {2006})}\BibitemShut {NoStop}%
\bibitem [{\citenamefont {Abraham}\ \emph
  {et~al.}(1995{\natexlab{a}})\citenamefont {Abraham}, \citenamefont
  {McAlexander}, \citenamefont {Sackett},\ and\ \citenamefont
  {Hulet}}]{abra95}%
  \BibitemOpen
  \bibfield  {author} {\bibinfo {author} {\bibfnamefont {E.~R.~I.}\
  \bibnamefont {Abraham}}, \bibinfo {author} {\bibfnamefont {W.~I.}\
  \bibnamefont {McAlexander}}, \bibinfo {author} {\bibfnamefont {C.~A.}\
  \bibnamefont {Sackett}},\ and\ \bibinfo {author} {\bibfnamefont {R.~G.}\
  \bibnamefont {Hulet}},\ }\bibfield  {title} {\bibinfo {title} {Spectroscopic
  determination of the s-wave scattering length of lithium},\ }\href@noop {}
  {\bibfield  {journal} {\bibinfo  {journal} {Phys. Rev. Lett.}\ }\textbf
  {\bibinfo {volume} {74}},\ \bibinfo {pages} {1315} (\bibinfo {year}
  {1995}{\natexlab{a}})}\BibitemShut {NoStop}%
\bibitem [{\citenamefont {Tsai}\ \emph {et~al.}(1997)\citenamefont {Tsai},
  \citenamefont {Freeland}, \citenamefont {Vogels}, \citenamefont {Boesten},
  \citenamefont {Verhaar},\ and\ \citenamefont {Heinzen}}]{tsai97}%
  \BibitemOpen
  \bibfield  {author} {\bibinfo {author} {\bibfnamefont {C.~C.}\ \bibnamefont
  {Tsai}}, \bibinfo {author} {\bibfnamefont {R.~S.}\ \bibnamefont {Freeland}},
  \bibinfo {author} {\bibfnamefont {J.~M.}\ \bibnamefont {Vogels}}, \bibinfo
  {author} {\bibfnamefont {H.~M. J.~M.}\ \bibnamefont {Boesten}}, \bibinfo
  {author} {\bibfnamefont {B.~J.}\ \bibnamefont {Verhaar}},\ and\ \bibinfo
  {author} {\bibfnamefont {D.~J.}\ \bibnamefont {Heinzen}},\ }\bibfield
  {title} {\bibinfo {title} {Two-color photoassociation spectroscopy of ground
  state ${\mathrm{Rb}}_{2}$},\ }\href
  {https://doi.org/10.1103/PhysRevLett.79.1245} {\bibfield  {journal} {\bibinfo
   {journal} {Phys. Rev. Lett.}\ }\textbf {\bibinfo {volume} {79}},\ \bibinfo
  {pages} {1245} (\bibinfo {year} {1997})}\BibitemShut {NoStop}%
\bibitem [{\citenamefont {Dutta}\ \emph {et~al.}(2017)\citenamefont {Dutta},
  \citenamefont {P\'erez-R\'{\i}os}, \citenamefont {Elliott},\ and\
  \citenamefont {Chen}}]{dutt17}%
  \BibitemOpen
  \bibfield  {author} {\bibinfo {author} {\bibfnamefont {S.}~\bibnamefont
  {Dutta}}, \bibinfo {author} {\bibfnamefont {J.}~\bibnamefont
  {P\'erez-R\'{\i}os}}, \bibinfo {author} {\bibfnamefont {D.~S.}\ \bibnamefont
  {Elliott}},\ and\ \bibinfo {author} {\bibfnamefont {Y.~P.}\ \bibnamefont
  {Chen}},\ }\bibfield  {title} {\bibinfo {title} {Two-photon photoassociation
  spectroscopy of an ultracold heteronuclear molecule},\ }\href
  {https://doi.org/10.1103/PhysRevA.95.013405} {\bibfield  {journal} {\bibinfo
  {journal} {Phys. Rev. A}\ }\textbf {\bibinfo {volume} {95}},\ \bibinfo
  {pages} {013405} (\bibinfo {year} {2017})}\BibitemShut {NoStop}%
\bibitem [{\citenamefont {Jones}\ \emph {et~al.}(2006)\citenamefont {Jones},
  \citenamefont {Tiesinga}, \citenamefont {Lett},\ and\ \citenamefont
  {Julienne}}]{jone06}%
  \BibitemOpen
  \bibfield  {author} {\bibinfo {author} {\bibfnamefont {K.~M.}\ \bibnamefont
  {Jones}}, \bibinfo {author} {\bibfnamefont {E.}~\bibnamefont {Tiesinga}},
  \bibinfo {author} {\bibfnamefont {P.~D.}\ \bibnamefont {Lett}},\ and\
  \bibinfo {author} {\bibfnamefont {P.~S.}\ \bibnamefont {Julienne}},\
  }\bibfield  {title} {\bibinfo {title} {Ultracold photoassociation
  spectroscopy: Long-range molecules and atomic scattering},\ }\href
  {https://doi.org/10.1103/RevModPhys.78.483} {\bibfield  {journal} {\bibinfo
  {journal} {Rev. Mod. Phys.}\ }\textbf {\bibinfo {volume} {78}},\ \bibinfo
  {pages} {483} (\bibinfo {year} {2006})}\BibitemShut {NoStop}%
\bibitem [{Note1()}]{Note1}%
  \BibitemOpen
  \bibinfo {note} {We will present our spectroscopy of this excited molecular
  state, which extends our previous observations in~\cite {roy16PA}, in a
  future publication.}\BibitemShut {Stop}%
\bibitem [{\citenamefont {Roy}\ \emph {et~al.}(2017)\citenamefont {Roy},
  \citenamefont {Green}, \citenamefont {Bowler},\ and\ \citenamefont
  {Gupta}}]{roy17}%
  \BibitemOpen
  \bibfield  {author} {\bibinfo {author} {\bibfnamefont {R.}~\bibnamefont
  {Roy}}, \bibinfo {author} {\bibfnamefont {A.}~\bibnamefont {Green}}, \bibinfo
  {author} {\bibfnamefont {R.}~\bibnamefont {Bowler}},\ and\ \bibinfo {author}
  {\bibfnamefont {S.}~\bibnamefont {Gupta}},\ }\bibfield  {title} {\bibinfo
  {title} {Two-element mixture of Bose and Fermi superfluids},\ }\href
  {https://doi.org/10.1103/PhysRevLett.118.055301} {\bibfield  {journal}
  {\bibinfo  {journal} {Phys. Rev. Lett.}\ }\textbf {\bibinfo {volume} {118}},\
  \bibinfo {pages} {055301} (\bibinfo {year} {2017})}\BibitemShut {NoStop}%
\bibitem [{\citenamefont {Abraham}\ \emph
  {et~al.}(1995{\natexlab{b}})\citenamefont {Abraham}, \citenamefont {Ritchie},
  \citenamefont {I.},\ and\ \citenamefont {Hulet}}]{abra95PA}%
  \BibitemOpen
  \bibfield  {author} {\bibinfo {author} {\bibfnamefont {E.~R.~I.}\
  \bibnamefont {Abraham}}, \bibinfo {author} {\bibfnamefont {N.~W.~M.}\
  \bibnamefont {Ritchie}}, \bibinfo {author} {\bibfnamefont {M.~W.}\
  \bibnamefont {I.}},\ and\ \bibinfo {author} {\bibfnamefont {R.~G.}\
  \bibnamefont {Hulet}},\ }\bibfield  {title} {\bibinfo {title} {Photoassociative spectroscopy of long-range states of ultracold $^{6}$}Li$_2$ and $^{7}$Li$_2$,\
  }\href@noop {} {\bibfield  {journal} {\bibinfo  {journal} {J. Chem. Phys.}\
  }\textbf {\bibinfo {volume} {103}},\ \bibinfo {pages} {7773} (\bibinfo {year}
  {1995}{\natexlab{b}})}\BibitemShut {NoStop}%
\bibitem [{\citenamefont {Leroy}\ and\ \citenamefont
  {Bernstein}(2006)}]{lero70}%
  \BibitemOpen
  \bibfield  {author} {\bibinfo {author} {\bibfnamefont {R.}~\bibnamefont
  {Leroy}}\ and\ \bibinfo {author} {\bibfnamefont {R.~B.}\ \bibnamefont
  {Bernstein}},\ }\bibfield  {title} {\bibinfo {title} {Dissociation energy and
  long-range potential of diatomic molecules from vibrational spacings at
  higher levels},\ }\href@noop {} {\bibfield  {journal} {\bibinfo  {journal}
  {J. Chem. Phys.}\ }\textbf {\bibinfo {volume} {52}},\ \bibinfo {pages} {3869}
  (\bibinfo {year} {2006})}\BibitemShut {NoStop}%
\bibitem [{\citenamefont {Roy}\ \emph {et~al.}(2016{\natexlab{b}})\citenamefont
  {Roy}, \citenamefont {Shrestha}, \citenamefont {Green}, \citenamefont
  {Gupta}, \citenamefont {Li}, \citenamefont {Kotochigova}, \citenamefont
  {Petrov},\ and\ \citenamefont {Yuen}}]{Roy2016}%
  \BibitemOpen
  \bibfield  {author} {\bibinfo {author} {\bibfnamefont {R.}~\bibnamefont
  {Roy}}, \bibinfo {author} {\bibfnamefont {R.}~\bibnamefont {Shrestha}},
  \bibinfo {author} {\bibfnamefont {A.}~\bibnamefont {Green}}, \bibinfo
  {author} {\bibfnamefont {S.}~\bibnamefont {Gupta}}, \bibinfo {author}
  {\bibfnamefont {M.}~\bibnamefont {Li}}, \bibinfo {author} {\bibfnamefont
  {S.}~\bibnamefont {Kotochigova}}, \bibinfo {author} {\bibfnamefont
  {A.}~\bibnamefont {Petrov}},\ and\ \bibinfo {author} {\bibfnamefont {C.~H.}\
  \bibnamefont {Yuen}},\ }\bibfield  {title} {\bibinfo {title}
  {Photoassociative production of ultracold heteronuclear ${\mathrm{YbLi}}^{*}$
  molecules},\ }\href {https://doi.org/10.1103/PhysRevA.94.033413} {\bibfield
  {journal} {\bibinfo  {journal} {Phys. Rev. A}\ }\textbf {\bibinfo {volume}
  {94}},\ \bibinfo {pages} {033413} (\bibinfo {year}
  {2016}{\natexlab{b}})}\BibitemShut {NoStop}%
\bibitem [{Note2()}]{Note2}%
  \BibitemOpen
  \bibinfo {note} {Deeper bound states ($v=-7$ and lower) were inaccessible in
  this work due to the choice of laser diodes used for PA. Future work can
  expand on the current study by using laser diodes centered at longer
  wavelengths.}\BibitemShut {Stop}%
  \bibitem [{Note3()}]{Note3}%
  \BibitemOpen
  \bibinfo {note} {The frequency range for our search included the expected frequency of the $v=-1$, $l=2$ resonance but our signal to noise ratio was not sufficient to observe it, likely because of low Franck-Condon factors.}\BibitemShut {Stop}%
\bibitem [{\citenamefont {Tang}\ and\ \citenamefont {Toennies}(1984)}]{Tang84}%
  \BibitemOpen
  \bibfield  {author} {\bibinfo {author} {\bibfnamefont {K.~T.}\ \bibnamefont
  {Tang}}\ and\ \bibinfo {author} {\bibfnamefont {J.~P.}\ \bibnamefont
  {Toennies}},\ }\bibfield  {title} {\bibinfo {title} {An improved simple model
  for the van der Waals potential based on universal damping functions for the
  dispersion coefficients},\ }\href {https://doi.org/10.1063/1.447150}
  {\bibfield  {journal} {\bibinfo  {journal} {J. Chem. Phys.}\
  }\textbf {\bibinfo {volume} {80}},\ \bibinfo {pages} {3726} (\bibinfo {year}
  {1984})}\BibitemShut {NoStop}%
\bibitem [{\citenamefont {Zhang}\ \emph {et~al.}(2010)\citenamefont {Zhang},
  \citenamefont {Sadeghpour},\ and\ \citenamefont {Dalgarno}}]{Zhang2010}%
  \BibitemOpen
  \bibfield  {author} {\bibinfo {author} {\bibfnamefont {P.}~\bibnamefont
  {Zhang}}, \bibinfo {author} {\bibfnamefont {H.~R.}\ \bibnamefont
  {Sadeghpour}},\ and\ \bibinfo {author} {\bibfnamefont {A.}~\bibnamefont
  {Dalgarno}},\ }\bibfield  {title} {\bibinfo {title} {Structure and
  spectroscopy of ground and excited states of LiYb},\ }\href
  {https://doi.org/10.1063/1.3462245} {\bibfield  {journal} {\bibinfo
  {journal} {J. Chem. Phys.}\ }\textbf {\bibinfo {volume}
  {133}},\ \bibinfo {pages} {044306} (\bibinfo {year} {2010})}\BibitemShut
  {NoStop}%
\bibitem [{\citenamefont {Brue}\ and\ \citenamefont {Hutson}(2013)}]{Brue2013}%
  \BibitemOpen
  \bibfield  {author} {\bibinfo {author} {\bibfnamefont {D.~A.}\ \bibnamefont
  {Brue}}\ and\ \bibinfo {author} {\bibfnamefont {J.~M.}\ \bibnamefont
  {Hutson}},\ }\bibfield  {title} {\bibinfo {title} {Prospects of forming
  ultracold molecules in ${}^{2}\sigma$ states by magnetoassociation of
  alkali-metal atoms with Yb},\ }\href
  {https://doi.org/10.1103/PhysRevA.87.052709} {\bibfield  {journal} {\bibinfo
  {journal} {Phys. Rev. A}\ }\textbf {\bibinfo {volume} {87}},\ \bibinfo
  {pages} {052709} (\bibinfo {year} {2013})}\BibitemShut {NoStop}%
\bibitem [{\citenamefont {Safronova}\ \emph {et~al.}(2012)\citenamefont
  {Safronova}, \citenamefont {Porsev},\ and\ \citenamefont
  {Clark}}]{Safronova2012}%
  \BibitemOpen
  \bibfield  {author} {\bibinfo {author} {\bibfnamefont {M.~S.}\ \bibnamefont
  {Safronova}}, \bibinfo {author} {\bibfnamefont {S.~G.}\ \bibnamefont
  {Porsev}},\ and\ \bibinfo {author} {\bibfnamefont {C.~W.}\ \bibnamefont
  {Clark}},\ }\bibfield  {title} {\bibinfo {title} {Ytterbium in quantum gases
  and atomic clocks: van der Waals interactions and blackbody shifts},\ }\href
  {https://doi.org/10.1103/PhysRevLett.109.230802} {\bibfield  {journal}
  {\bibinfo  {journal} {Phys. Rev. Lett.}\ }\textbf {\bibinfo {volume} {109}},\
  \bibinfo {pages} {230802} (\bibinfo {year} {2012})}\BibitemShut {NoStop}%
\bibitem [{\citenamefont {Porsev}\ \emph {et~al.}(2014)\citenamefont {Porsev},
  \citenamefont {Safronova}, \citenamefont {Derevianko},\ and\ \citenamefont
  {Clark}}]{Porsev2014}%
  \BibitemOpen
  \bibfield  {author} {\bibinfo {author} {\bibfnamefont {S.~G.}\ \bibnamefont
  {Porsev}}, \bibinfo {author} {\bibfnamefont {M.~S.}\ \bibnamefont
  {Safronova}}, \bibinfo {author} {\bibfnamefont {A.}~\bibnamefont
  {Derevianko}},\ and\ \bibinfo {author} {\bibfnamefont {C.~W.}\ \bibnamefont
  {Clark}},\ }\bibfield  {title} {\bibinfo {title} {Relativistic many-body
  calculations of van der Waals coefficients for Yb-Li and Yb-Rb dimers},\
  }\href {https://doi.org/10.1103/PhysRevA.89.022703} {\bibfield  {journal}
  {\bibinfo  {journal} {Phys. Rev. A}\ }\textbf {\bibinfo {volume} {89}},\
  \bibinfo {pages} {022703} (\bibinfo {year} {2014})}\BibitemShut {NoStop}%
\bibitem [{\citenamefont {Thakkar}\ and\ \citenamefont
  {Smith}(1974)}]{Thakkar1974}%
  \BibitemOpen
  \bibfield  {author} {\bibinfo {author} {\bibfnamefont {A.~J.}\ \bibnamefont
  {Thakkar}}\ and\ \bibinfo {author} {\bibfnamefont {V.~H.}\ \bibnamefont
  {Smith}},\ }\bibfield  {title} {\bibinfo {title} {Mist: A new interatomic
  potential function},\ }\href
  {https://doi.org/https://doi.org/10.1016/0009-2614(74)85423-0} {\bibfield
  {journal} {\bibinfo  {journal} {Chem. Phys. Lett.}\ }\textbf {\bibinfo
  {volume} {24}},\ \bibinfo {pages} {157 } (\bibinfo {year}
  {1974})}\BibitemShut {NoStop}%
\bibitem [{\citenamefont {Gopakumar}\ \emph {et~al.}(2010)\citenamefont
  {Gopakumar}, \citenamefont {Abe}, \citenamefont {Das}, \citenamefont {Hada},\
  and\ \citenamefont {Hirao}}]{Gopakumar2010}%
  \BibitemOpen
  \bibfield  {author} {\bibinfo {author} {\bibfnamefont {G.}~\bibnamefont
  {Gopakumar}}, \bibinfo {author} {\bibfnamefont {M.}~\bibnamefont {Abe}},
  \bibinfo {author} {\bibfnamefont {B.~P.}\ \bibnamefont {Das}}, \bibinfo
  {author} {\bibfnamefont {M.}~\bibnamefont {Hada}},\ and\ \bibinfo {author}
  {\bibfnamefont {K.}~\bibnamefont {Hirao}},\ }\bibfield  {title} {\bibinfo
  {title} {Relativistic calculations of ground and excited states of LiYb
  molecule for ultracold photoassociation spectroscopy studies},\ }\href
  {https://doi.org/10.1063/1.3475568} {\bibfield  {journal} {\bibinfo
  {journal} {J. Chem. Phys.}\ }\textbf {\bibinfo {volume}
  {133}},\ \bibinfo {pages} {124317} (\bibinfo {year} {2010})}\BibitemShut
  {NoStop}%
\bibitem [{\citenamefont {Kotochigova}\ \emph {et~al.}(2011)\citenamefont
  {Kotochigova}, \citenamefont {Petrov}, \citenamefont {Linnik}, \citenamefont
  {Kłos},\ and\ \citenamefont {Julienne}}]{Kotochigova2011}%
  \BibitemOpen
  \bibfield  {author} {\bibinfo {author} {\bibfnamefont {S.}~\bibnamefont
  {Kotochigova}}, \bibinfo {author} {\bibfnamefont {A.}~\bibnamefont {Petrov}},
  \bibinfo {author} {\bibfnamefont {M.}~\bibnamefont {Linnik}}, \bibinfo
  {author} {\bibfnamefont {J.}~\bibnamefont {Kłos}},\ and\ \bibinfo {author}
  {\bibfnamefont {P.~S.}\ \bibnamefont {Julienne}},\ }\bibfield  {title}
  {\bibinfo {title} {Ab initio properties of Li-group-II molecules for
  ultracold matter studies},\ }\href {https://doi.org/10.1063/1.3653974}
  {\bibfield  {journal} {\bibinfo  {journal} {J. Chem. Phys.}\
  }\textbf {\bibinfo {volume} {135}},\ \bibinfo {pages} {164108} (\bibinfo
  {year} {2011})}\BibitemShut {NoStop}%
\bibitem [{\citenamefont {Tohme}\ \emph {et~al.}(2015)\citenamefont {Tohme},
  \citenamefont {Korek},\ and\ \citenamefont {Awad}}]{Tohme2015}%
  \BibitemOpen
  \bibfield  {author} {\bibinfo {author} {\bibfnamefont {S.~N.}\ \bibnamefont
  {Tohme}}, \bibinfo {author} {\bibfnamefont {M.}~\bibnamefont {Korek}},\ and\
  \bibinfo {author} {\bibfnamefont {R.}~\bibnamefont {Awad}},\ }\bibfield
  {title} {\bibinfo {title} {Ab initio calculations of the electronic structure
  of the low-lying states for the ultracold LiYb molecule},\ }\href
  {https://doi.org/10.1063/1.4914472} {\bibfield  {journal} {\bibinfo
  {journal} {J. Chem. Phys.}\ }\textbf {\bibinfo {volume}
  {142}},\ \bibinfo {pages} {114312} (\bibinfo {year} {2015})}\BibitemShut
  {NoStop}%
\bibitem [{Note4()}]{Note4}%
  \BibitemOpen
  \bibinfo {note} {We neglect small effects such as couplings between the
  rotational angular momentum of the diatom and the spins, including the
  nuclear and the electronic ones, and the dependence on the nuclear separation
  $R$ of the hyperfine constant and the {\protect \it g}-factors. Thus the
  spins are completely decoupled from the nuclear motion, and the hyperfine
  splittings of the molecular bound states in a magnetic field is the same as
  the ones in the separate-atoms limit.}\BibitemShut {Stop}%
\bibitem [{\citenamefont {Mohr}\ and\ \citenamefont
  {Taylor}(2000)}]{CODATA1998}%
  \BibitemOpen
  \bibfield  {author} {\bibinfo {author} {\bibfnamefont {P.~J.}\ \bibnamefont
  {Mohr}}\ and\ \bibinfo {author} {\bibfnamefont {B.~N.}\ \bibnamefont
  {Taylor}},\ }\bibfield  {title} {\bibinfo {title} {CODATA recommended values
  of the fundamental physical constants: 1998},\ }\href
  {https://doi.org/10.1103/RevModPhys.72.351} {\bibfield  {journal} {\bibinfo
  {journal} {Rev. Mod. Phys.}\ }\textbf {\bibinfo {volume} {72}},\ \bibinfo
  {pages} {351} (\bibinfo {year} {2000})}\BibitemShut {NoStop}%
\bibitem [{Note5()}]{Note5}%
  \BibitemOpen
  \bibinfo {note} {Although $l_v$ are used as unknown variables, they are
  discrete and remain zero in the vicinity of the minimum where the error
  propagation is practically done. Therefore we don't count them as variables
  in the error analysis.}\BibitemShut {Stop}%
\bibitem [{\citenamefont {Gordon}(1969)}]{Gordon69}%
  \BibitemOpen
  \bibfield  {author} {\bibinfo {author} {\bibfnamefont {R.~G.}\ \bibnamefont
  {Gordon}},\ }\bibfield  {title} {\bibinfo {title} {New method for
  constructing wavefunctions for bound states and scattering},\ }\href
  {https://doi.org/10.1063/1.1671699} {\bibfield  {journal} {\bibinfo
  {journal} {J. Chem. Phys.}\ }\textbf {\bibinfo {volume} {51}},\ \bibinfo
  {pages} {14} (\bibinfo {year} {1969})}\BibitemShut {NoStop}%
\bibitem [{\citenamefont {Ivanov}\ \emph {et~al.}(2011)\citenamefont {Ivanov},
  \citenamefont {Khramov}, \citenamefont {Hansen}, \citenamefont {Dowd},
  \citenamefont {M\"unchow}, \citenamefont {Jamison},\ and\ \citenamefont
  {Gupta}}]{Ivanov2011}%
  \BibitemOpen
  \bibfield  {author} {\bibinfo {author} {\bibfnamefont {V.~V.}\ \bibnamefont
  {Ivanov}}, \bibinfo {author} {\bibfnamefont {A.}~\bibnamefont {Khramov}},
  \bibinfo {author} {\bibfnamefont {A.~H.}\ \bibnamefont {Hansen}}, \bibinfo
  {author} {\bibfnamefont {W.~H.}\ \bibnamefont {Dowd}}, \bibinfo {author}
  {\bibfnamefont {F.}~\bibnamefont {M\"unchow}}, \bibinfo {author}
  {\bibfnamefont {A.~O.}\ \bibnamefont {Jamison}},\ and\ \bibinfo {author}
  {\bibfnamefont {S.}~\bibnamefont {Gupta}},\ }\bibfield  {title} {\bibinfo
  {title} {Sympathetic cooling in an optically trapped mixture of alkali and
  spin-singlet atoms},\ }\href {https://doi.org/10.1103/PhysRevLett.106.153201}
  {\bibfield  {journal} {\bibinfo  {journal} {Phys. Rev. Lett.}\ }\textbf
  {\bibinfo {volume} {106}},\ \bibinfo {pages} {153201} (\bibinfo {year}
  {2011})}\BibitemShut {NoStop}%
\bibitem [{\citenamefont {Hara}\ \emph {et~al.}(2011)\citenamefont {Hara},
  \citenamefont {Takasu}, \citenamefont {Yamaoka}, \citenamefont {Doyle},\ and\
  \citenamefont {Takahashi}}]{Hara2011}%
  \BibitemOpen
  \bibfield  {author} {\bibinfo {author} {\bibfnamefont {H.}~\bibnamefont
  {Hara}}, \bibinfo {author} {\bibfnamefont {Y.}~\bibnamefont {Takasu}},
  \bibinfo {author} {\bibfnamefont {Y.}~\bibnamefont {Yamaoka}}, \bibinfo
  {author} {\bibfnamefont {J.~M.}\ \bibnamefont {Doyle}},\ and\ \bibinfo
  {author} {\bibfnamefont {Y.}~\bibnamefont {Takahashi}},\ }\bibfield  {title}
  {\bibinfo {title} {Quantum degenerate mixtures of alkali and
  alkaline-earth-like atoms},\ }\href
  {https://doi.org/10.1103/PhysRevLett.106.205304} {\bibfield  {journal}
  {\bibinfo  {journal} {Phys. Rev. Lett.}\ }\textbf {\bibinfo {volume} {106}},\
  \bibinfo {pages} {205304} (\bibinfo {year} {2011})}\BibitemShut {NoStop}%
\bibitem [{\citenamefont {Gao}(2008)}]{Gao2008}%
  \BibitemOpen
  \bibfield  {author} {\bibinfo {author} {\bibfnamefont {B.}~\bibnamefont
  {Gao}},\ }\bibfield  {title} {\bibinfo {title} {General form of the
  quantum-defect theory for $-\nicefrac{1}{r^\alpha}$ type of potentials with $\alpha > 2$},\ }\href
  {https://doi.org/10.1103/PhysRevA.78.012702} {\bibfield  {journal} {\bibinfo
  {journal} {Phys. Rev. A}\ }\textbf {\bibinfo {volume} {78}},\ \bibinfo
  {pages} {012702} (\bibinfo {year} {2008})}\BibitemShut {NoStop}%
\bibitem [{\citenamefont {Harris}(1997)}]{harr97}%
  \BibitemOpen
  \bibfield  {author} {\bibinfo {author} {\bibfnamefont {S.~E.}\ \bibnamefont
  {Harris}},\ }\bibfield  {title} {\bibinfo {title} {Electromagnetically
  induced transparency},\ }\href@noop {} {\bibfield  {journal} {\bibinfo
  {journal} {Physics Today}\ }\textbf {\bibinfo {volume} {50}},\ \bibinfo
  {pages} {36} (\bibinfo {year} {1997})}\BibitemShut {NoStop}%
\bibitem [{\citenamefont {Winkler}\ \emph {et~al.}(2005)\citenamefont
  {Winkler}, \citenamefont {Thalhammer}, \citenamefont {Theis}, \citenamefont
  {Ritsch}, \citenamefont {Grimm},\ and\ \citenamefont
  {Denschlag}}]{wink05}%
  \BibitemOpen
  \bibfield  {author} {\bibinfo {author} {\bibfnamefont {K.}~\bibnamefont
  {Winkler}}, \bibinfo {author} {\bibfnamefont {G.}~\bibnamefont {Thalhammer}},
  \bibinfo {author} {\bibfnamefont {M.}~\bibnamefont {Theis}}, \bibinfo
  {author} {\bibfnamefont {H.}~\bibnamefont {Ritsch}}, \bibinfo {author}
  {\bibfnamefont {R.}~\bibnamefont {Grimm}},\ and\ \bibinfo {author}
  {\bibfnamefont {J.~H.}~\bibnamefont {Denschlag}},\ }\bibfield  {title}
  {\bibinfo {title} {Atom-molecule dark states in a Bose-Einstein condensate},\
  }\href@noop {} {\bibfield  {journal} {\bibinfo  {journal} {Phys. Rev. Lett.}\
  }\textbf {\bibinfo {volume} {95}},\ \bibinfo {pages} {063202} (\bibinfo
  {year} {2005})}\BibitemShut {NoStop}%
\bibitem [{\citenamefont {Stellmer}\ \emph {et~al.}(2012)\citenamefont
  {Stellmer}, \citenamefont {Pasquiou}, \citenamefont {Grimm},\ and\
  \citenamefont {Schreck}}]{stel12}%
  \BibitemOpen
  \bibfield  {author} {\bibinfo {author} {\bibfnamefont {S.}~\bibnamefont
  {Stellmer}}, \bibinfo {author} {\bibfnamefont {B.}~\bibnamefont {Pasquiou}},
  \bibinfo {author} {\bibfnamefont {R.}~\bibnamefont {Grimm}},\ and\ \bibinfo
  {author} {\bibfnamefont {F.}~\bibnamefont {Schreck}},\ }\bibfield  {title}
  {\bibinfo {title} {Creation of ultracold Sr$_2$ molecules in the electronic
  ground state},\ }\href@noop {} {\bibfield  {journal} {\bibinfo  {journal}
  {Phys. Rev. Lett.}\ }\textbf {\bibinfo {volume} {109}},\ \bibinfo {pages}
  {115302} (\bibinfo {year} {2012})}\BibitemShut {NoStop}%
\bibitem [{\citenamefont {Lu}\ and\ \citenamefont {Wang}(2017)}]{luwa17}%
  \BibitemOpen
  \bibfield  {author} {\bibinfo {author} {\bibfnamefont {B.}~\bibnamefont
  {Lu}}\ and\ \bibinfo {author} {\bibfnamefont {D.}~\bibnamefont {Wang}},\
  }\bibfield  {title} {\bibinfo {title} {Note: A four-pass acousto-optic
  modulator system for laser cooling of sodium atoms},\ }\href@noop {}
  {\bibfield  {journal} {\bibinfo  {journal} {Rev. Sci. Instr.}\ }\textbf
  {\bibinfo {volume} {88}},\ \bibinfo {pages} {076105} (\bibinfo {year}
  {2017})}\BibitemShut {NoStop}%
\bibitem [{\citenamefont {Ciamei}\ \emph {et~al.}(2017)\citenamefont {Ciamei},
  \citenamefont {Bayerle}, \citenamefont {Chen}, \citenamefont {Pasquiou},\
  and\ \citenamefont {Schreck}}]{ciam17}%
  \BibitemOpen
  \bibfield  {author} {\bibinfo {author} {\bibfnamefont {A.}~\bibnamefont
  {Ciamei}}, \bibinfo {author} {\bibfnamefont {A.}~\bibnamefont {Bayerle}},
  \bibinfo {author} {\bibfnamefont {C.~C.}~\bibnamefont {Chen}}, \bibinfo {author}
  {\bibfnamefont {B.}~\bibnamefont {Pasquiou}},\ and\ \bibinfo {author}
  {\bibfnamefont {F.}~\bibnamefont {Schreck}},\ }\bibfield  {title} {\bibinfo
  {title} {Efficient production of long-lived ultracold Sr$_2$ molecules},\
  }\href@noop {} {\bibfield  {journal} {\bibinfo  {journal} {Phys. Rev. A.}\
  }\textbf {\bibinfo {volume} {96}},\ \bibinfo {pages} {013406} (\bibinfo
  {year} {2017})}\BibitemShut {NoStop}%
\bibitem [{Note6()}]{Note6}%
  \BibitemOpen
  \bibinfo {note} {We have recently observed magnetic Feshbach resonances in
  Yb-Li at locations consistent with predictions based on the work presented in
  this paper. These observations will be reported in a forthcoming
  publication.}\BibitemShut {Stop}%
\bibitem [{\citenamefont {Mark}\ \emph {et~al.}(2018)\citenamefont {Mark},
  \citenamefont {Meinert}, \citenamefont {Lauber},\ and\ \citenamefont
  {N$\ddot{\rm a}$gerl}}]{mark18}%
  \BibitemOpen
  \bibfield  {author} {\bibinfo {author} {\bibfnamefont {M.}~\bibnamefont
  {Mark}}, \bibinfo {author} {\bibfnamefont {F.}~\bibnamefont {Meinert}},
  \bibinfo {author} {\bibfnamefont {K.}~\bibnamefont {Lauber}},\ and\ \bibinfo
  {author} {\bibfnamefont {H.-C.}~\bibnamefont {N$\ddot{\rm a}$gerl}},\ }\bibfield
   {title} {\bibinfo {title} {Mott-insulator-aided detection of ultra-narrow
  Feshbach resonances},\ }\href@noop {} {\bibfield  {journal} {\bibinfo
  {journal} {SciPost Phys.}\ }\textbf {\bibinfo {volume} {5}},\ \bibinfo
  {pages} {055} (\bibinfo {year} {2018})}\BibitemShut {NoStop}%
\end{thebibliography}
\end{document}